\documentstyle[eqsecnum,aps,preprint]{revtex}
\def\vev#1{\left\langle #1\right\rangle}

\def\beq {\begin{equation}}
\def\eeq {\end{equation}}
\def \a  {\alpha}
\def \be {\beta}
\def \cE {{\cal E}}
\def \cL {{\cal L}}
\def \cM {{\cal M}}
\def \cN {{\cal N}}
\def \cI {{\cal I}}
\def \cJ {{\cal J}}
\def \cH {{\cal H}}
\def \cR {{\cal R}}
\def \cT {{\cal T}}
\def \cA {{\cal A}}
\def \cB {{\cal B}}
\def \cP {{\cal P}}
\def \cF {{\cal F}}
\def \half {{1\over 2}}
\def\lr#1{\vbox{\baselineskip 1.5pt \hbox{$\leftrightarrow$}\par\hbox{$#1$}}}
\def\CC{\hbox{C\kern -.53em {\raise .4ex \hbox{$\scriptscriptstyle |$}}
 \kern-.51em {\raise .5ex \hbox{$\scriptscriptstyle |$}} }}
\def\HH{\hbox{I\kern-.21em\hbox{H}}}
\def\RR{\hbox{I\kern-.2em\hbox{R}}}
\def\sRR{{\sl \hbox{I\kern-.2em\hbox{R}}}}
\def\ZZ{{{\rm Z}\kern-.28em{\rm Z}}}
\def\sqr#1#2{{\vcenter{\vbox{\hrule height.#2pt\hbox{\vrule width.#2pt
height#1pt \kern#1pt\vrule width.#2pt}\hrule height.#2pt}}}}
\def\square{\mathchoice\sqr54\sqr54\sqr33\sqr23}

\begin{document}
\draft
\preprint{WISC-MILW-94-TH-22}
\title { Existence and uniqueness theorems for massless fields\\
on a class of spacetimes with closed timelike curves.}

\author {John L. Friedman}

\address{Institute for Theoretical Physics\\
       University of California\\
       Santa Barbara, CA  93106\\
\centerline {and}
 Department of Physics\\
       University of Wisconsin-Milwaukee\\
       Milwaukee, WI   53201}
\author {Michael S. Morris}
\address{ Department of Physics \& Astronomy\\
       Butler University\\
       4600 Sunset Ave. \\
       Indianapolis, IN 46208}
\date{\today}
\maketitle
\begin{abstract}
We study the massless scalar field on asymptotically flat spacetimes
with closed timelike curves (CTC's), in which all future-directed CTC's
traverse one end of a handle (wormhole) and emerge from the other end
at an earlier time.  For a class of static geometries of this type, and
for smooth initial data with all derivatives in $L_2$ on ${\cI}^{-}$,
we prove existence of smooth solutions which are regular at null and
spatial infinity (have finite energy and finite $L_2$-norm) and have
the given initial data on $\cI^-$.   A restricted uniqueness theorem is
obtained, applying to solutions that fall off in time at any fixed
spatial position.  For a complementary class of spacetimes in which
CTC's are confined to a compact region, we show that when solutions
exist they are unique in regions exterior to the CTC's.  (We believe
that more stringent uniqueness theorems hold, and that the present
limitations are our own.)  An extension of these results to Maxwell
fields and massless spinor fields is sketched.  Finally, we discuss a
conjecture that the Cauchy problem for free fields is well defined in
the presence of CTC's whenever the problem is well-posed in the
geometric-optics limit.  We provide some evidence in support of this
conjecture, and we present counterexamples that show that neither
existence nor uniqueness is guaranteed under weaker conditions.  In
particular, both existence and uniqueness can fail in smooth,
asymptotically flat spacetimes with a compact nonchronal region.
\end{abstract}

\pacs{}
\narrowtext

\section{Introduction}
\label{sec:intro}

Although spacetimes with closed timelike curves occur as solutions to
the vacuum Einstein equations, they have been regarded as unphysical, in
part because the most familiar examples have no well-defined initial
value problem.   Morris and Thorne \cite{mt}, however, introduced a
class of wormhole geometries in which, although there are many closed
timelike curves, the set of closed timelike and null geodesics has
measure zero. On these spacetimes, Morris {\it et.  al.}\cite{mty}
noted that the evolution of free fields is well defined in the limit of
geometrical optics; and this in turn makes it seem likely that a
multiple scattering series converges to a solution for arbitrary
initial data.\cite{mty,fmetal,fm91}  The simplest of the spacetimes
they considered are static, with CTC's present at all times, and we
reported an existence and a restricted uniqueness theorem for the
massless scalar field on such static time-tunnel spacetimes.\cite{fm91}

The present paper provides details of these latter results and outlines their
extension to spinor and vector fields.  A final section discusses the
Cauchy problem on spacetimes in which CTCs are confined to a compact
region.  We prove a uniqueness theorem and present a conjecture on the
existence of free fields whose data are specified on a spacelike
hypersurface (partial Cauchy surface) to the past of any CTCs.

Until the final section, the geometry $\cN, g$ we consider is static
in the sense that there is a timelike Killing vector $t^\alpha$ that is
everywhere locally hypersurface orthogonal.\footnote
{Spacetime indices will be lower case Greek, spatial indices lower case
Latin, spinor indices upper case Latin.  For those familiar with the
abstract index notation \cite{wald}, letters near a (or $\alpha$) in
the alphabet will be abstract while those after $i$ ($\iota$) will be
concrete, so that $\xi^\mu$ is the $\mu^{th}$ component of the vector
$\xi^\alpha$.  Our signature is $-+++$, and Riemann tensor conventions
follow \cite{wald}.  We will use ``manifold'' as a shorthand for smooth
manifold, with or without boundary.}
The manifold has topology ${\cal N} = M\times \sRR$, where $M$ is a
plane with a handle (wormhole) attached: \pagebreak
$M = \sRR^3\#(S^2\times S^1)$.
The metric $g_{\alpha\beta}$ on ${\cal N}$ is smooth
($C^\infty$), and, for simplicity in treating  the asymptotic behavior
of the fields, we will assume that outside a compact region ${\cal R}$
the geometry is flat, with metric $\eta_{\alpha\beta}$.

\section{Preliminaries}
\label{sec:prelim}

\subsection{Foliation of a related spacetime with boundary.}
\label{sec:spacetime}

One can construct the 3-manifold $M$ from $\sRR^3$ by removing two
balls and identifying their spherical boundaries, $\Sigma_I$ and
$\cT(\Sigma_I)$, by a map $\cT$, as shown in Fig.~\ref{M}.  In
Fig.~\ref{M}, the map $\cT$ involves an improper rotation of
$\Sigma_I$, and it yields an orientable handle.  Maps $\cT$ which
involve a proper rotation of $\Sigma_I$ yield non-orientable handles.
For the rest of this paper, we will in fact assume that $\cT$ is
improper and that the handle is consequently orientable. This matters,
however, only to our treatment of a two-component Weyl spinor.  Were we
to allow non-orientable handles as well, everything would be the same,
except that we should have to consider 4-component Dirac spinors.  The
sphere obtained by the identification of $\Sigma_I$ and $\cT(\Sigma_I)$
will be called the ``throat'' of the handle.  Its location is
arbitrary:  after removing any sphere $\Sigma$, from the handle of $M$
one is left with a manifold homeomorphic to $\sRR^3\backslash(B^3\#
B^3)$, whose boundary is the disjoint union of two spheres.  Let $C
\cong S^2\times \sRR$ be the history of the throat in the spacetime
${\cal N}$, its orbit under the group of time-translations generated by
$t^\alpha$.  Then, after removing the cylinder $C$ from the spacetime
${\cal N}$, one is similarly left with a manifold homeomorphic to
$\sRR^4\backslash[(B^3\# B^3)\times\sRR]$, whose boundary, $\partial
({\cal N}\backslash C)$, is the disjoint union, $C_I\bigsqcup C_{II}$,
of two timelike cylinders.  We will again use the symbol $\cT$ to
denote the map from $C_I$ to $C_{II}$ that relates identified points;
its restriction to a single sphere $\Sigma_I\subset C_I$ is the map
denoted above by $\cT$.

For the spacetimes we will consider, identified points of $C_I$ and
$C_{II} = \cT(C_I)$
will be timelike separated, and copies of $M$ in
the spacetime cannot be everywhere spacelike.  Thus, although
$t^\alpha$ is locally hypersurface-orthogonal, no complete
hypersurface of $\cN$ with a single asymptotically flat region
is orthogonal to every trajectory of
$t^\alpha$.   One can, however, foliate $\cN\backslash C$ by spacelike
hypersurfaces $\cM_t$ orthogonal to $t^\alpha$.  Each $\cM_t$ can be
chosen to agree asymptotically with a $t$ = constant surface of the
flat metric $\eta_{\alpha\beta}$ and to intersect $C_I$ and $C_{II}$ in
isometric spheres $\Sigma_I$ and $\Sigma_{II}$.   The sphere
$\cT(\Sigma_I)$ identified with $\Sigma_I$ is a time-translation of
$\Sigma_{II}$, obtained by moving $\Sigma_{II}$ a parameter distance
$\tau$ along the trajectories of $t^\alpha$ (see Fig.~\ref{Mt}).

The manifold $\cM_t$ near $\Sigma_I$ is a smooth extension of
$\cM_{t+\tau}$ near $\cT(\Sigma_I)$.  That is, if $U_I\subset\cM_t$ and
$U_{t+\tau}\subset\cM_{t+\tau}$ are spacelike neighborhoods of
$\Sigma_I$ and $\cT(\Sigma_I)$, their union $U_I\cup U_{t+\tau}$ is a
smooth, spacelike submanifold of $\cN$.  Because the time-translation
of $U_{t+\tau}$ to $\cM_t$ is a neighborhood $U_{II}\subset\cM_t$ of
$\Sigma_{II}$ isometric to $U_{t+\tau}$, by {\it artificially}
identifying the spheres $\Sigma_I$ and $\Sigma_{II}$ of $\cM$, one
obtains a copy of $M$ (a plane with a handle) with a metric that is
everywhere smooth and spacelike.  Of course this spacelike copy of $M$ is
not a submanifold of the spacetime $\cN$.

A static metric on $\cN$ is given by
\begin{equation}
 g_{\alpha\beta} = - e^{-2\nu}t_\alpha t_\beta + h_{\alpha\beta}, \label{gab1}
\end{equation}
where $h_{\alpha\beta} t^\beta = 0$.  If the Minkowski coordinate $t$
is extended to $\cN\backslash C$ by making $\cM_t$ a $t$=constant
surface, then $t^\alpha \nabla_\alpha t = 1$, $\nabla^\alpha t = -
e^{-2\nu}t^\alpha$, and the metric (\ref{gab1}) can be written on
$\cN\backslash C$ in the form
\begin{equation}
g_{\alpha\beta} = - e^{2\nu}\partial_\alpha t \partial_\beta t
+ h_{\alpha\beta}.\label{gab2}\end{equation}
It will be convenient to single out a representative hypersurface,
\begin{equation}
\cM := \cM_0. \label{cm}
\end{equation}
We will denote by $h_{ab}$ the corresponding spatial metric on $\cM$;
that is, $h_{ab}$ is the pullback of $h_{\alpha\beta}$ (or
$g_{\alpha\beta}$) to $\cM$.

\subsection{An initial value problem}
\label{sec:ivalue}

Although ${\cal N}$ has no spacelike hypersurface which could
play the role of a Cauchy surface, one can pose initial data
to the massless scalar wave equation
\begin{equation}
\sqr75\ \Phi = 0 \label{scwaveeq}
\end{equation}
at past null infinity, $\cI^-$. We will show that for all data in
$H_\infty(\cI^-)$, there is a solution $\Phi$ to Eq.~(\ref{scwaveeq}).

Because the geometry is flat outside a region of fixed
spatial size, ${\cal I}$ is a copy of the Minkowski space ${\cal I}$.
In the null chart $(v=t+r,\vec{r})$, $\cI^-$ has coordinates $(v,
\hat{r})$. In Minkowski space, a smooth solution $\Phi_0$ with finite
energy to the wave equation (\ref{scwaveeq})
has as initial data on ${\cal I}^{-}$ the single function \cite{wilcox}
\begin{equation}
f_0(v,\hat{r}) = \lim_{r \rightarrow \infty}
r\Phi_0(v,r\hat{r}). \label{id1}
\end{equation}
It is helpful to write the solution $\Phi_0$
in terms of its positive frequency part, $\Psi_0$:
\begin{equation}
\Phi_0 = 2 {\rm Re} \Psi_0 = (2\pi)^{-3/2}\int d^3 k\ [a(k) e^{i(k\cdot x
-\omega t)} + a^*(k) e^{-i(k\cdot x -\omega t)}].\label{phi0}
\end{equation}
Then the function,
\begin{equation}
\Psi_0 =  (2\pi)^{-3/2}\int d^3 k\ a(k) e^{i(k\cdot x -\omega t)}\label{psi0},
\end{equation}
has initial data on $\cI^-$ given by \cite{wilcox}
\begin{equation}
\lim_{r \rightarrow \infty} r\Psi_0(v,r\hat r)
= {i\over (2\pi)^\half}\int_0^\infty d\omega\
\omega\ a(-\omega\hat r) e^{-i\omega v}.\label{id2}
\end{equation}
The corresponding initial data $f_0(v,\hat r)$ for $\Phi_0$,
has Fourier transform,
\begin{equation}
\tilde f_0 (\omega,\hat r) = {1 \over (2\pi)^\half} \int_{-\infty}^\infty dv
f_0 (v,\hat r) e^{i\omega v},\label{tf}
\end{equation}
related to $a(k)$ by
\begin{equation}
\tilde f_0 (\omega,\hat r) = i\omega a(-\omega \hat r),\quad \omega\geq 0,
\label{fa1}
\end{equation}
\begin{equation}
\tilde f_0 (-\omega, \hat r) = \tilde f_0 ^*(\omega, \hat r).\label{fa2}
\end{equation}

If we define $L_2(\cI^-)$ by the norm,
\begin{equation}
\left|\left|f\right|\right|^2_{L_2(\cI^-)} = \int dv d\Omega |f|^2
\label{l2scri},
\end{equation}
then the $L_2$ norm of $\Psi_0$ on a spacelike hyperplane is equal to
the $L_2$ norm of its initial data on $\cI^-$:
\begin{equation}
\lim_{r \rightarrow \infty} \int dv d\Omega\
|r\Psi_0|^2 = \int dk\ |a|^2 = \int\ dV\ |\Psi_0|^2.  \label{l2psi}
\end{equation}
The $L_2$ norm of $\Phi_0$ depends on hypersurface, but it is
bounded by the constant $L_2$ norm of $\Psi_0$.

The flux of energy at $\cI^-$ is given in terms of the stress tensor,
\begin{equation}
T_{\alpha\beta} = \nabla_\alpha \Phi \nabla_\beta \Phi -\
\half\ g_{\alpha\beta}\nabla_\gamma\Phi\nabla^\gamma\Phi,\label{stab}
\end{equation}
by
\begin{eqnarray}\int_{\cI^-} dS_\alpha\ T^\alpha_\beta\ t^\beta
&=& \lim_{r \rightarrow \infty}\int dv d\Omega\ [\partial_v(r\Phi_0)]^2
\label{sflux}\\
&=&\int d\omega d\Omega\ \omega^2|\tilde f|^2\\
&=&\int dk\ \omega^2|a(k)|^2.
\end{eqnarray}

In the spacetime $\cN$ that we are considering, we seek a solution
$\Phi$ to the scalar wave equation,
\begin{equation}
\nabla_\alpha \nabla^\alpha \Phi = 0, \label{kg}
\end{equation}
with initial data $f$ for which $f$ and its derivatives are in
$L_2(\cI^-)$ .  It will again be convenient to relate $f$ to a function
$a(k)$ as in Eqs. (\ref{tf}) - (\ref{fa2}), in this case defining
$a(k)$ by Eqs. (\ref{fa1}) and (\ref{fa2}).

Initial data for vector and spinor fields have a similar
character.  An electromagnetic field $F_{\alpha\beta}$ can be written in terms
of a vector potential ${\rm A}_\alpha$ satisfying the Lorentz gauge condition,
\begin{equation}
F_{\alpha\beta} =  \nabla_\alpha {\rm A}_\beta - \nabla_\beta
{\rm A}_\alpha,\qquad\nabla_\alpha {\rm A}^\alpha = 0.\label{lor}
\end{equation}
The equation $\nabla_\beta F_\alpha^\beta = 0$,
governing a free Maxwell field, is then equivalent to
\begin{equation}
\nabla_\beta \nabla^\beta {\rm A}^\alpha - R^\alpha_\beta {\rm A}^\beta = 0.
\label{max}
\end{equation}
For a field with finite energy on Minkowski space, initial data
on $\cI^-$ has the form,
\begin{equation}
\lim_{r\rightarrow\infty}r{\rm A}_\alpha(v,r\hat r) = \ 2 {\rm Re}{i\over
(2\pi)^\half}\int d\omega\ \omega a_\alpha(-\omega \hat r)\ e^{-i\omega v},
\label{idem}
\end{equation}
with $a_\alpha(k) k^\alpha = 0 = a_\alpha t^\alpha$.

We will adopt the 2-component spinor notation given in Penrose and
Rindler \cite{pr}, with $\nabla_{AA'} = \sigma^\alpha_{AA'}\nabla_\alpha$,
where $ \sigma^\alpha_{AA'}$ has components equal to entries of (the usual
Pauli spin matrices)$/\sqrt2$.  The free-field equation for a massless
spinor $\Phi^A$ is given by
\begin{equation}
\nabla_{AA'}\Phi^A = 0,\label{weyl}\end{equation}
with initial data on $\cI^-$
\begin{equation}
\lim_{r\rightarrow\infty}r\Phi^A(v,r\hat r) =\ 2 {\rm Re}{i\over
(2\pi)^\half} \int d\omega\ \omega a^A(-\omega \hat r)\ e^{-i\omega
v},\label{idsp}\end{equation}
where $a^A(k) k_{AA'} = 0.$
Because the spacetime $\cN$ is not simply connected, one
must specify a choice of spinor structure in order make sense of Eq.
(\ref{weyl}).  On the simply connected spacetime $\cN\backslash C$, the two
spinor structures on $\cN$ correspond to a choice of sign in
identifying a spinor at a point $P_I\in C_I$  with a spinor at the
corresponding point $P_{II}\in C_{II}$.  The choice of spinor structure
thus becomes a choice of boundary condition (see Eq. (\ref{wbc}), below).
The two choices give two inequivalent spinor fields on $\cN$.

As in (\ref{l2psi}), the $L_2$ norm of the initial data for vector and
spinor fields is equal in flat space to the $L_2$ norm on a spacelike
hyperplane:
\begin{equation}
\lim_{r \rightarrow \infty} \int dv d\Omega\  |r {\rm A}_{0}|^2 = \int
dk\ |a|^2 = \int dV \ |{\rm A}|^2 ,
\label{l2v}
\end{equation}
\begin{equation}
\lim_{r \rightarrow \infty} \int dv d\Omega\  |r \Phi_{0}|^2 = \int
dk\ |a|^2 = \int dV \ \ |\Phi|^2.
\label{l2sp}
\end{equation}
{}From the form of the energy-momentum tensor for vector and spinor fields,
\begin{equation}
T_{\alpha\beta} = {1\over 4\pi} (F_{\alpha\gamma} F_\beta^{\ \gamma} -
{1\over 4} g_{\alpha\beta} F_{\gamma\delta}F^{\gamma\delta}),
\label{maxtab}
\end{equation}
\begin{equation}
T_{\alpha\beta} = i
\sigma_\alpha^{AA'}\sigma_\beta^{BB'}(\Phi_{(A}\nabla_{B)A'}
\bar\Phi_{B'} - \bar\Phi_{(A'}\nabla_{B')A} \Phi_{B}),
\label{wtab}
\end{equation}
the energy flux of the fields $F_{\alpha\beta}$ and $\Phi^A$ at $\cI^-$ is
\begin{equation}
{1\over 4\pi}\int_{\cI^-} dS_\alpha\ F^{\alpha\gamma}F_{\beta\gamma}\ t^\beta =
{1\over 4\pi} \int dk\ \omega^2\ a_j(k)^*a^j(k),
\label{maxflux}
\end{equation}
and
\begin{equation}
\int_{\cI^-} dS_\alpha\ \sigma^{aAA'}\sigma_\beta^{BB'}(\Phi_{(A}\nabla_{B)A'}
\bar\Phi_{B'} - \bar\Phi_{(A'}\nabla_{B')A} \Phi_{B})\ t^\beta =
\int dk\ \omega^2\ |a(k)|^2.
\label{wflux}
\end{equation}

\subsection{Boundary conditions}
\label{sec:bc}

A scalar field on $\cN$ satisfies at the cylindrical boundaries of
$\cN\backslash C$ conditions expressing the continuity of $\Phi$ and
its normal derivative along a path that traverses the wormhole.  Let
$P_I$ and $P_{II}= \cT(P_I)$ be identified points on the cylinders
$C_I$ and $C_{II}$.
The tangent vector to a path that traverses the wormhole, entering at
$P_I$ and leaving at $P_{II}$, points inward at $P_I$ and outward at
$P_{II}$.
When the cylinders $C_I$ and $C_{II}$ are identified, a unit inward normal
to $C_I$ is thus identified with a unit outward normal to $C_{II}$.  If
we denote by $\hat n_I$ and $\hat n_{II}$ the unit outward normals to $C_I$
and $C_{II}$, the boundary conditions can be written
\begin{mathletters}
\label{kgbc}
\begin{equation}
\Phi(P_{II}) = \Phi(P_I)\label{kgbca}\end{equation}
\begin{equation}
\hat n_{II}\cdot \nabla\Phi(P_{II}) =
-\hat n_I\cdot\nabla\Phi(P_I).
\label{kgbcb}
\end{equation}
\end{mathletters}

The analogous conditions for vector and spinor fields can be stated in
terms of the differential map $\cT_*$ induced by $\cT$.
If $\{\hat e_\mu(P_I)\} = \{\hat e_0, \hat e_1, \hat e_2, n_I\}$
is a right-handed  orthonormal frame at $P_I$, then the corresponding
right-handed frame at $P_{II}$ (if $\cT$ were a proper rotation, then
the corresponding frame would be left-handed) is
\begin{equation}
\{\hat e_\mu(P_{II})\} = \{\cT_*\hat e_0, \cT_*\hat e_1,
\cT_*\hat e_2,\  -\hat n_{II}\}.
\label{basis}
\end{equation}
The boundary conditions for a vector field can be expressed
in terms of its components along the frame $\{\hat e_\mu\}$:
\begin{mathletters}
\label{vbc}
\begin{equation}
{\rm A}_\mu(P_{II}) = {\rm A}_\mu(P_I)\label{vbca}
\end{equation}
\begin{equation}
 \hat n_{II}\cdot \nabla {\rm A}_\mu(P_{II}) =
-\hat n_I\cdot\nabla {\rm A}_\mu(P_I). \label{vbcb}
\end{equation}
\end{mathletters}

A spinor field has components along a spinor frame, an element of the
double covering ($\simeq SL(2,$\CC) ) of the space of right-handed
orthonormal frames ($\simeq$ SO(3,1))
at a point.  Two spinor frames
correspond to the same orthonormal frame, and the choice of spinor
structure is the choice of which spinor frame at $P_{II}$ to identify
with a spinor frame at $P_I$.  Let $(o^A, \iota^A)$ be a field of
spinor frames covering a field of frames $\hat e_\mu$
that satisfies Eq.~(\ref{basis}) on $\cN\backslash C$.  We can choose as
boundary conditions for the corresponding components $\Phi^J$ of a
spinor field
\begin{mathletters}
\label{wbc}
\begin{equation}
\Phi^J(P_{II}) =  \Phi^J(P_I)
\label{wbca}
\end{equation}
\begin{equation}
\hat n_{II}\cdot \nabla\Phi^J(P_{II}) =
-\hat n_I\cdot\nabla\Phi^J(P_I).
\label{wbcb}
\end{equation}
\end{mathletters}
The opposite spinor structure would be selected by changing the sign of
the right hand side of these equations or, equivalently, by keeping the
same sign but choosing a homotopically different frame field $\{\vec
e_\mu\}$.

\subsection{ Eigenfunction expansions}
\label{sec:efn}

Because the geometry is static, we can express
solutions as a superposition of functions with harmonic time
dependence.
\begin{mathletters}
\label{harm}
\begin{equation}
\Phi(t,x) = \int d\omega \phi(\omega,x) e^{-i\omega t},
\end{equation}
\begin{equation}
{\rm A}_\alpha(t,x) = \int d\omega A_\alpha(\omega,x) e^{-i\omega t},
\end{equation}
\begin{equation}
\Phi^A(t,x) = \int d\omega \phi^A(\omega,x) e^{-i\omega t}.
\end{equation}
\end{mathletters}
Here $x$ is naturally a point of the manifold of trajectories of $t^\alpha$,
but we can identify it with a point of the simply connected spacelike
hypersurface $\cM = \cM_0$, with spherical boundaries $\Sigma_I$ and
$\Sigma_{II}$.  Let $(t, x_I)$ and $(t+\tau, x_{II})$ be
points of $\cN\backslash C$ that are identified in $\cN$.
The harmonic components of
fields on $\cN$ can be regarded as fields on $\cM$ satisfying the
boundary conditions,
\begin{mathletters}
\label{kghb}
\begin{equation}
\phi(\omega,x_{II}) =  e^{i\eta} \phi(\omega,x_I),
\label{kghba}
\end{equation}
\begin{equation}
\hat n_{II}\cdot \nabla\phi(\omega,x_{II}) =
- e^{i\eta} \hat n_I\cdot\nabla\phi(\omega, x_I),
\label{kghbb}
\end{equation}
\end{mathletters}
\begin{mathletters}
\label{vhb}
\begin{equation}
{\rm A}_\mu(\omega, x_{II}) =  e^{i\eta}{\rm A}_\mu( \omega, x_I )
\label{vhba}
\end{equation}
\begin{equation}
\hat n_{II}\cdot \nabla {\rm A}_\mu(\omega, x_{II}) =
- e^{i\eta} \hat n_I\cdot\nabla {\rm A}_\mu( \omega, x_I ),
\label{vhbb}
\end{equation}
\end{mathletters}
\begin{mathletters}
\label{whb}
\begin{equation}
\Phi^J(\omega, x_{II}) =   e^{i\eta} \Phi^I(\omega, x_I ),
\label{whba}
\end{equation}
\begin{equation}
\hat n_{II}\cdot \nabla\phi^J(\omega, x_{II}) =
- e^{i\eta} \hat n_I\cdot\nabla\phi^J(\omega, x_I),
\label{whbb}
\end{equation}
\end{mathletters}
with phase $\eta = \omega\tau$.

The harmonic components, $\phi$, of the scalar field satisfy on $\cM$ elliptic
equations of the form
\begin{equation}
(\omega^2 + \cL)\phi = 0,
\label{lphi}
\end{equation}
with boundary conditions (\ref{kghb}), where $\cL$ can be defined by
the action of $\nabla_\alpha \nabla^\alpha$ on time independent
fields $f$ on $\cN$; that is
\begin{equation}
\cL f := e^{2\nu}\nabla_\alpha \nabla^\alpha f |_\cM,
\label{defl}
\end{equation}
for fields satisfying $\mbox{\pounds}_t f =0$, where $\mbox{\pounds}_t $ is
the Lie derivative along $ t^{\alpha} $.
Then
\begin{equation}
\cL = e^{\nu}D^a e^\nu D_a, \label{defl2}
\end{equation}
where $D_a$ is the covariant derivative of the 3-metric $h_{ab}$ on $\cM$.
We will denote by $\cL_\eta$ the operator $\cL$ with boundary
conditions (\ref{kghb}).  The analogous operators for spinor and
scalar fields are discussed in Sect.~\ref{sec:other}.
We show in {\sl Lemma 4} below that one can construct solutions
$F(\eta, k, x)$, to the wave
equation (\ref{kg}), having, on the flat geometry outside
${\cal R}$, the form of a plane wave plus a purely outgoing wave,
corresponding to the scattering of the plane wave by the interior
geometry.   The existence of solutions for initial data on $\cI^-$ then
follows if one can show that a spectral decomposition of the form
\begin{equation}
\Phi (x,t) = \int E(k,x) \Bigl[ {\rm e}^{-i\omega t}\,a(k) + {\rm
e}^{i\omega t}\,a^\ast(k)\Bigr]d^3 k,\label{sspectral}\end{equation}
converges, where $E(k,x) = F(\eta = \omega\tau, k, x)$.  The major
difficulty lies in the fact that, because the boundary conditions
(\ref{kgbc}) involve a time-translation, the corresponding boundary
conditions (\ref{kghb}) depend on the frequency $\omega$.  If $\eta$ were
independent of frequency, the result would follow from the usual
spectral theorem for self-adjoint operators.  Here, however, $E(k,x)$ and
$E(k^\prime,x)$ are, for $|k| \neq |k^\prime|$, eigenfunctions of
different operators; they are not orthogonal, and their completeness is
not guaranteed by the spectral theorem.  The main job of Section III is
to show that the solution to the scalar wave equation for arbitrary
initial data on ${\cal I}^{-}$ \ can nevertheless be constructed as a
spectral integral of the form (\ref{sspectral}).

We adopt throughout the common usage in which the term
``eigenfunction'' refers not to an element of the domain of
${\cL_\eta}$, but to a function in a weighted $L_2$ space, whose
$L_2$ norm diverges (an example is the function $ e^{ikx}$ in $\sRR^3$).

\subsection{Sobolev spaces}
\label{sec:sobolev}

We shall need some standard properties of Sobolev spaces, including the
Sobolev embedding and trace theorems.  These can be found in Reed and
Simon \cite{reedsimon}.  Denote by $H_{s}(N)$ the Sobolev space on a
manifold $N$ with volume form $\epsilon$, so that for $s$ a positive
integer, $H_{s}(N)$ is the space of functions on $N$ for which
the function and its first $s$ derivatives are square integrable.
More generally, for any real $s$ and any chart $y:U\rightarrow {\sl \sRR^n}$
of $N$, we have:
\hfill\break {\sl Definition}. The Hilbert space $H_{s}(U)$ is the
completion of $C_0^\infty(U)$ in the norm
\begin{equation}
\| f \|_s = \int_{\sl \sRR^n} d\xi\ |\hat f (\xi) |^2\ (1+|\xi|^2)^s,
\label{hsloc}
\end{equation}
where $\hat f$ is the Fourier transform on ${\sl \sRR^n}$ of $f\circ y^{-1}$,
regarded as a function on ${\sl \sRR^n }$ with support on $y(U)$.

We use several spaces of functions based on the Sobolev spaces $H_s(N)$.\\
{\sl Definition}.  $H_{s}^{loc}(N)$ is the space of functions in
$H_{s}(U)$ for all compact $U\subset N$.  \hfill\break
For the spacelike
3-manifold $\cM \equiv \cM_0$ a single chart $x:\cM\rightarrow
{\sl \sRR }^3$ will map the flat metric outside some region ${\cal R}$
to the flat metric
of ${\sl \sRR }^3$, allowing a simple definition of Sobolev spaces and
weighted $L_2$ spaces on $\cM$:\hfill\break
{\sl Definition}.  $H_{s}(\cM)$ is the completion of $C_0^\infty(\cM)$
in the norm (\ref{hsloc}), with $\hat f$ the Fourier transform of
$f\circ x^{-1}$.   $L_{2,r}(\cM)$ is the completion of
$C_0^\infty(\cM)$ in the norm
\begin{equation}
\| f \|_{2,r} =\int_\cM dV e^{-\nu} |f|^2 (1+|x|^2)^r.\label{l2r}\end{equation}
The lapse function, $e^{-\nu}$, appearing in the measure is required to
make the operator $\cL$ symmetric.\\ The spaces $H_s({\sl \sRR}^3)$ and
$L_{2,r}({\sl \sRR}^3)$ are defined in the same way, with $\cM$
replaced by ${\sl \sRR}^3$.

A tensor field will be said to be in any of the spaces defined above if
its components with respect to the charts $y$ or $x$ are in the space.
A spinor field will be said to be in any of these spaces if its
components in a spinor frame are in the spaces, for a spinor frame
associated with a smooth orthonormal frame that has bounded covariant
derivatives of all orders.

Finally, it is helpful to define spaces that incorporate the boundary
conditions on \\
$\partial \cM = \Sigma_I\bigsqcup\Sigma_{II}$.\hfill\break
{\sl Definition}.  $\HH_1$ is the intersection of $H_1(\cM)$ with
functions satisfying the boundary condition (\ref{kgbca}).  $\HH_2$ is
the intersection of $H_2(\cM)$ with functions satisfying the boundary
conditions (\ref{kgbca}-\ref{kgbcb}). \hfill\break
The Sobolev trace theorem implies that elements of $H_1(\cM)$ have
well-defined values on $\partial \cM$, so $\HH_1$ and $\HH_2$ are
completions in the $H_1$ and $H_2$ norms of $C^\infty$ functions
satisfying the boundary conditions specified in the definition.

\section{Existence and uniqueness theorems}
\label{sec:existunique}

\subsection{Existence theorem for a massless scalar field}
\label{sec:exist}

Let $R$ be large enough that for $r>R$ the geometry is flat.  Define an
exterior region ${\cal E}$ by ${\cal E} = \{p\in \cN |\ r(p) > R\}$ and
an interior region $\cR = \cN\backslash{\cal E}$.  It will be helpful in what
follows to introduce a smoothed step function, $\chi$, that vanishes
on $\cR$:
\begin{equation}
\chi(p)  = \cases{0,&if $p\in \cR$\cr
                  1,&if $r(p)\geq R+\epsilon$,}
\label{chi}
\end{equation}
for some $\epsilon>0$.\\
{\sl Definition}.  A scalar field $\Phi$ on $\cN$ is {\sl
asymptotically regular at spatial infinity} if $\Phi\circ\chi\in
H_1(\cM)$; it is {\sl asymptotically regular at null infinity} if the limits,
\begin{eqnarray}
f(v,\hat r) = \lim_{r \rightarrow \infty} r\Phi(v,r\hat{r}),\nonumber\\
g(u,\hat r) = \lim_{r \rightarrow \infty} r\Phi(u,r\hat{r}),
\end{eqnarray}
exist, with $f\in H_1(\cI^-),\ g\in H_1(\cI^+)$, where $u$ is the null
coordinate $t-r$.\\
That is, $\Phi$ is asymptotically regular if it is an $L_2$-function
with finite energy on $\cN$ and has well-defined data with finite
energy on $\cI^-$ and $\cI^+$.

Our result on existence of solutions to the scalar wave equation has
the following form.  \\
{\sl Proposition 1}.  For almost all spacetimes ${\cal N}, g$ of the
kind described in Sec.~\ref{sec:spacetime} (for almost all parameters
$\tau$), the following existence theorem holds. Let $f$ be initial data
on ${\cal I}^{-}$ for which f and all its derivatives are in $L_2({\cal
I}^{-})$.  Then there exists a solution $\Phi$ to the scalar wave
equation which is smooth and asymptotically regular at null and spatial
infinity and which has $f$ as initial data.

The proof is given as a sequence of lemmas.  For boundary conditions
(\ref{kghb}) specified by a fixed phase, $\eta$, we show that
the operators $\cL_\eta$ are self-adjoint on a dense subspace of
$L_2$.   We follow a
method given by Wilcox \cite{wilcox} to obtain explicit eigenfunctions
$F(\eta, k, x)$ in a weighted $L_2$ space.  For fixed boundary phase $\eta$
the eigenfunctions are complete and orthonormal, but, as mentioned earlier,
a solution $\Phi$ of Eq.~(\ref{kg}) is a superposition of
eigenfunctions,
\begin{equation}
E(k, x) =\ F(\eta=\omega\tau,\ k,\ x),
\label{ekx}
\end{equation}
that are not orthonormal.  For each $\omega$, the function $E(k, x)$ is an
eigenfunction of a different operator $\cL_\eta$ because the boundary
condition depends on $\omega$ through the relation $\eta = \omega\tau$.

\noindent
{\sl Lemma 1}.  The operator $\cL_{\eta}$ with boundary conditions
(\ref{kghb}) is self-adjoint on the space $L_2(\cal M)$
with domain $\HH_2$.

\noindent{\sl Proof.}
Recall that for an operator $A$ with domain $D(A) \subset L_2(\cM)$, a
function $f\in L_2(\cM)$ is in $D(A^\dagger)$ if and only
if $\exists h\in L_2(\cM)$ such that $<f|\ Ag>\ =\ <h|\ g> \forall g\in D(A)$.
To
prove that $\HH_2\subset D(\cL^\dagger_{\eta})$ is easy:  for $f\in\HH_2$,
$\cL_{\eta} f\in L_2(\cM)$.  Take $h=\cL_{\eta} f$.  Then, writing
$\Sigma:=\Sigma_I\sqcup\Sigma_{II}$, we have
\begin{equation}
<f|\ \cL_\eta g> = <\cL_\eta f|\ g> + \int_{\Sigma_I\sqcup\Sigma_{II}}
dS_a e^{-\nu}\ (\overline {f e^{i\eta}})
\lr D^a g e^{i\eta}.
\label{sa1}
\end{equation}
Because $f$ and $g$ satisfy the boundary conditions (\ref{kghb}),
we have
\begin{equation}
\int_{\Sigma_{II}} dS_a e^{-\nu}\ \bar f \lr D^a g =
-\int_{\Sigma_I} dS_a e^{-\nu}\ (\overline {f e^{i\eta}})
\lr D^a g e^{i\eta} = -\int_{\Sigma_I} dS_a e^{-\nu}\ \bar
f \lr D^a g.
\label{sa2}
\end{equation}
Thus $\cL_{\eta}$ is symmetric,
\begin{equation}
\langle f\mid\cL_{\eta} g\rangle = \langle\cL_{\eta} f\mid g \rangle=
\langle h\mid g
\rangle,
\label{sa3}
\end{equation}
and $\HH_2\subset D(\cL^\dagger_{\eta} )$.

To prove that $D(\cL^\dagger_{\eta} )\subset \HH_2$, we proceed as follows.
For $f\in D(\cL^\dagger_{\eta})$, there is an $h\in L_2(\cM)$ with
\begin{equation}
<f|\ \cL_{\eta} g>\ =\ <h|\ g> \forall g\in D(\cL_{\eta}).
\label{sa4}\end{equation}

That is, the equation $\cL^\dagger_{\eta} f = h$ is satisfied weakly on $\cM$.
Then, by elliptic regularity, we have $f\in H_2^{\rm loc}(\cM)$ and
$\cL_{\eta} f = h$ is satisfied strongly on $\cM$.

Asymptotic behavior of $f$ follows from the fact that the exterior region
${\cal E}$ can be regarded as a subspace of Euclidean 3-space, $E^3$.
Let $\chi$ be the smooth step function of Eq.~(\ref{chi}), and let
$\tilde f = f\chi$.  Then $\tilde f$ is a function on $E^3$ satisfying
$\nabla^2 \tilde f = \tilde h$, with $h\in L_2(E^3)$.  Because
$\nabla^2$ is self-adjoint on $E^3$ with domain $H_2(E^3)$, we have
$\tilde f\in H_2(E^3)$, implying $f|{\cal E}\in H_2({\cal E})$.

To check that $f$ satisfies the boundary conditions, we use the fact,
noted in Sect. I (before Eq.~(\ref{gab1})), that by identifying
$\Sigma_I$ and $\Sigma_{II}$, one obtains a spacelike copy $\hat M$ of
$M$ with a smooth spatial metric, $\hat g_{ab}$. ($\hat M$ is an
artificially constructed spacelike surface with a handle; it is not a
submanifold of our spacetime, $\cN$.) The operator $\hat\cL$
constructed from $\hat g_{ab}$ is smooth and elliptic.  Because the
projection, $p:  \cM_t\rightarrow M$, is an isometry there,
it agrees with $\cL$ on the interior of $\cM_t$.  Let $U$
be a neighborhood of $\Sigma$ for which $p^{-1}(U) = U_I\sqcup U_{II}$,
with $U_I$ and $U_{II}$ disjoint neighborhoods of $\Sigma_I$ and
$\Sigma_{II}$.  If, as expected, the function $f\circ p^{-1}$ has a phase
that changes by $e^{i\eta}$ across $\Sigma$, then we should get a
smooth function $\hat f$ on  $U$ by requiring
\begin{eqnarray}
\hat f(P) & = &\left\{
\begin{array}{l}
	\;\; f\circ p^{-1},\; P\in p(U_I)\\
 	e^{-i\eta}f\circ p^{-1},\; P\in p(U_{II})
\end{array}\right.
\label{sa5}
\end{eqnarray}
To see that this is true, first smoothly truncate $f$ so that it
vanishes outside of $U_{1}$ and $U_{II}$, and then define $\hat f$ as
above.  Let $g$ of (\ref{sa4}) similarly have support on $U_{1}\cup
U_{II}$, and let $h$ be the function in $L_2(M)$ given by (\ref{sa5}).
Define $\hat g$ and $\hat h$  on $U_{1}\cup
U_{II}$ according to (\ref{sa5}), with $f$
replaced by $g$ and $h$, respectively.  Then $g\in \HH_2
(\cM)\Rightarrow\hat g\in H_2(\hat M)$, and we have \begin{equation}
<\hat f|\ \cL_{\eta} \hat g>\ =\ <\hat h|\ \hat g> ,
\label{sa6}
\end{equation}
$\forall \hat g\in H_2(\hat M)$ with support on $U_{1}\cup U_{II}$,
implying (again by elliptic regularity) that $\hat f$ is smooth across
$\Sigma$.  Thus any $f\in D(\cL^\dagger_{\eta} )$ is a function in
$H_2$ satisfying the boundary conditions (\ref{kghb}), meaning
$D(\cL^\dagger_{\eta} )\subset\HH_2$.$\qquad\qquad\square$

We now find a set of eigenfunctions that are complete for data on
${\cal I}^-$. We consider solutions $F(\eta,k,x)$ to Eq.~(\ref{lphi})
which, for $r>R$, have the form
\begin{equation}
F =(2\pi)^{-3/2}e^{ik\cdot x}+{\rm outgoing\ waves}.
\label{fk}
\end{equation}

To prove existence of the eigenfunctions $F$, one
first rewrites Eq.~(\ref{lphi}) in the inhomogeneous form
\begin{equation}
(\omega^2 + \cL_{\eta}) \varphi=\rho
\label{inhom}
\end{equation}
where $\varphi$ is purely outgoing
and $\rho$ has compact support. One can do
this by using the steplike function $\chi(r)$ of Eq.~(\ref{chi}), writing
\begin{equation}
F =\ \chi(2\pi)^{-3/2} e^{ik\cdot x}+F_{\rm out}.
\label{fout}
\end{equation}
Then
\begin{equation}
(\omega^2 + \cL_{\eta}) F=(\omega^2+ \cL_{\eta} )F_{\rm out}-\rho,
\label{kinhom}
\end{equation}
with
\begin{equation}
\rho = -(2\pi)^{-3/2}e^{ik\cdot x}
(\nabla^\beta\nabla_\beta-2ik^\beta\nabla_\beta)\chi.
\label{rho}
\end{equation}
The homogeneous equation, $(\omega^2 + \cL_{\eta}) F=0$, is equivalent
to the inhomogeneous equation (\ref{inhom}) with $\varphi = F_{\rm
out}$.\\

\noindent{\sl Lemma 2}. Let $\lambda_k$ be a sequence of complex
numbers with positive imaginary part, such that
$\lambda_k\to\omega^2$.  Consider families $\{\varphi_k,\ \rho_k\}$
of smooth fields on $\cM$, where for each $k$, the field $\rho_k$ has
support on the region $r\leq R+\epsilon$, and $\varphi_k$ is the unique
asymptotically regular solution to Eq. (\ref{inhom}) with
$\omega^2$ replaced by $\lambda_k$.  If $\rho_k\to\rho$ in $L_2(\cM)$, then a
subsequence $\{\varphi_m\}$ converges in an $H_1$ norm to a smooth
outgoing solution $\varphi$ to Eq.~(\ref{inhom}).
\vspace{0.2in}

\noindent
{\sl Proof}. Let $D$ be a real number greater than $R+\epsilon$,
$\cM_D=\{x\in\cM,\ r\leq D\}$. Denote by $\|\ \|_{s,D}$ the norm of
$H_s(\cM_D)$.

Suppose first that $\|\varphi_k\|_{2,D}$
has a bound $C$ independent of $k$. Since $H_2(\cM_D)\hookrightarrow
H_1(\cM_D)$ is a compact embedding (by the Sobolev embedding theorem) the
set $\{\varphi_k\}$ belongs to a compact set of
$H_1(\cM_D)$. Thus there is a subsequence
$\{\varphi_m\}$ that converges in $H_1(\cM_D)$:
\begin{equation}
\varphi_m\to\varphi. \label{limphi}
\end{equation}

We must show that $\varphi$ satisfies (i) $ ( \omega^2 + \cL ) \varphi = \rho $
on $\cM_D$,
(ii) the boundary conditions (\ref{kghb}), and (iii) we
must extend $\varphi$ outside $\cM_D$.\hfill\break
(i): To see that $\varphi$ weakly satisfies (\ref{inhom}), define
$\forall\psi\in C^\infty_0 (\cM_D)$
\begin{equation}
Q\equiv-\vev{D_a (e^\nu\psi)|D^a (e^\nu\varphi)} +\omega^2\vev{\psi|\varphi} -
\vev{\psi|\rho}. \label{Q}
\end{equation}
Add to this $Q$ to the quantity
$\vev{D_a (e^\nu\psi)|D^a (e^\nu\varphi_m)} -\lambda_m\vev{\psi|\varphi_m}
+\vev{\psi|\rho_m}=0$. Then
\begin{eqnarray}
|Q| & =
|-\vev{D_a (e^\nu\psi)|D^a (e^\nu\varphi)}+\omega^2\vev{\psi|\varphi_m}
-\vev{\psi|\rho} \nonumber\\
& + \vev{D_a (e^\nu\psi)|D^a (e^\nu\varphi_m)}
-\lambda_m\vev{\psi|\varphi_m} + \vev{\psi|\rho_m}|\nonumber\\
& \leq\left|\vev{D_a (e^\nu\psi) |D^a [e^\nu(\varphi-\varphi_m)]}\right|
+\omega^2\left|\vev{\psi|\varphi-\varphi_m}\right|
+ \nonumber\\
& \left|(\omega^2-\lambda_m)\vev{\psi|\varphi_m}\right|
+\left|\vev{\psi|\rho-\rho_m}\right|.\label{absq}
\end{eqnarray}
The limit of the right hand side of (\ref{absq}) is zero as $m\to\infty$:
\begin{mathletters}\label{limphi2}\begin{eqnarray}
\lim_{m\to\infty}\|\varphi-\varphi_m\|_{1,R} &= 0,\label{limphia}\\
\lim_{m\to\infty}\|\rho-\rho_m\|_0\  &= 0,\label{limphib}\\
\lim_{m\to\infty}|\omega^2-\lambda_m|\  &= 0.\label{limphic}
\end{eqnarray}\end{mathletters}
Then $Q$ independent of $m$ implies $Q=0$.  Hence $\varphi$ is a weak
solution to $ ( \omega^2 + \cL ) \varphi = \rho $
in $\cM_D$, and elliptic regularity implies
that it is a strong solution.\hfill\break

\noindent
(ii): Next define a neighbourhood $U$ on $\hat M $, as in the proof of
{\sl Lemma 1}. And define $ { \hat \varphi}_k$ on $U$ from
${\varphi_k}$ on $ \cM_D$ in exactly the same way that $ \hat f $ was
defined in {\sl Lemma 1} from $f$. Then, because $ {\varphi_k} $ is
smooth and satisfies the boundary conditions (\ref{kghb}) on $\cM_D$,
we have that $ {\hat \varphi}_k $ is smooth on $U$. Also, because
$\rho_k$ has support only outside of $U$ (away from the boundaries), $
{\hat \varphi}_k $ satisfies $ ( \lambda_k
+ \hat \cL ) {\hat \varphi}_k = 0 $ on $U$.  Now, a subsequence
$ \{ {\hat \varphi}_m \} $ converges in $H_1 ( U ) $ to a solution
$ {\hat \varphi} $, which is continuous. But, by elliptic regularity
$ {\hat \varphi} $ is smooth in $U$ and therefore $ \varphi$ satisfies both
boundary conditions (\ref{kghb})
$\forall\rho\in C^\infty(\cM_D)$. \hfill\break
(iii): For $r>R+\epsilon$, we have $\rho_m=0$. Because the space is
flat outside $r=R$, and $\varphi_m$ is asymptotically regular, we can
use the explicit Green function for $\nabla^2_{{\rm flat}}+\omega^2$ to
write
\begin{equation}
\varphi_m(x)=\int_{|y|=R}dS_y\  \varphi_m(y)\ \lr{\partial_y}\
{e^{i\sqrt{\lambda_m}|x-y|}\over|x-y|},\label{phim}
\end{equation}
where ${\rm Im}\sqrt{\lambda_m}>0$. Then for $R+\epsilon<r<D$,
\begin{equation}
\varphi(x)=\lim_{m\to\infty}\varphi_m(x)=\int_{|y|=R} dS_y\
\varphi(y)\ \lr{\partial_y}\ {e^{i\omega|x-y|}\over|x-y|},
\label{outgoing}
\end{equation}
an outgoing wave. Defining $\varphi$ by Eq.~(\ref{outgoing}) for $r>D$
we obtain an outgoing $C^\infty$ solution to (\ref{inhom}) as claimed.

The construction has so far relied on the assumption that
$\|\varphi_k\|_{2,D}$, was bounded. If not, the sequence
$\tilde\varphi_k=\varphi_k/\|\varphi_k\|_{2,D}$ has unit norm and a
source $\tilde\rho_k=\rho_k/\|\varphi_k\|_{2,D}$ whose norm converges
to zero:
\begin{equation}
\lim_{k\to\infty}\|\tilde\rho_k\|=0. \label{limrho}
\end{equation}
This leads to a contradiction. From the previous paragraph, there is a
subsequence $\tilde\phi_m$ converging to an outgoing solution $\tilde\varphi$
of $\cL_\eta\tilde\varphi=0$. But Rellich's uniqueness theorem ({\sl Lemma~3}
below) implies $\tilde\varphi=0$, whence
$\lim_{m\to\infty}\|\varphi_m\|_{0,D}=0$. From
$\cL\tilde\varphi_m=\tilde\rho_m$, we have $\|\tilde\varphi_m\|_{2,D}\leq
C\|\tilde\varphi_m\|_{0,D}+\|\rho_m\|_0$. Then
$\lim_{m\to\infty}\|\tilde\varphi_m\|_{2,D}=0$ contradicting
$\|\tilde\varphi_m\|_{2,D}=1$.\\

\noindent{\sl Lemma 3}. (Rellich's uniqueness theorem.) Let $\varphi\in
\HH_2$ be outgoing at spatial infinity and satisfy $(\cL_\eta +
\omega^2) \varphi=0$.
Then $\varphi=0$.

\noindent{\sl Proof.} In the exterior region ${\cal E}$, $\varphi$ is a
smooth solution to the flat space equation
$(\omega^2+\nabla^2)\varphi=0$ and can therefore be written as a sum
\begin{equation}
\sum\a_{lm}h^{(1)}_l(\omega r)Y_{l
m}(\Omega)+\be_{lm}h_l^{(2)}(wr)Y_{l m}(\Omega),
\label{hlm1}\end{equation}
converging in $L_2^{{\rm loc}}$.   Here $h^{(1)}_l$ and
$h^{(2)}_l=h^{(1)*}_l$ are spherical Hankel functions, satisfying
the Wronskian relation,
\begin{equation}
h^{(1)}_l(y)\partial_y h^{(2)}_l(y) - h^{(2)}_l(y)\partial_y h^{(1)}_l(y)
={2\over{i y^2}}
\label{wronskian}\end{equation}
Using this relation and orthonormality of the spherical harmonics
$Y_{l m}$, we have,
\begin{eqnarray}
0 & = & \int_\cR dV \varphi^* (\omega^2+\cL) \varphi
    - \int_{\cR} dV (\omega^2+\cL)\varphi^* \varphi
=\int_{r=R} dS\ \varphi^*\ \lr{\partial_r}\ \varphi\nonumber\\
 & = & \ \sum_{lm} \Big[|\a_{lm}|^2
 h^{(2)}_l\ \lr{\partial_r}\ h^{(1)}_l+|\be_{lm}|^2h^{(1)}_l
\ \lr{\partial_r}\ h^{(2)}_l\Big] R^2\nonumber\\
 & = & {2i\over\omega}\sum[|\a_{lm}|^2-|\be_{lm}|^2]\label{hlm2}\\
\Longrightarrow\     \sum|\a_{lm}|^2 & = & \sum|\be_{lm}|^2.
\label{hlm3}\end{eqnarray}
In other words, ingoing and outgoing fluxes are equal. Then $\varphi$
outgoing implies
$\a_{lm}=0\Longrightarrow\ \be_{lm}=0\Longrightarrow\     \varphi=0$
outside $\cR$.  Aronszjan's elliptic continuation theorem
\cite{aronzjan} then implies $\varphi=0$ everywhere on $\cM$. $\square$

\vspace{0.2in}
\noindent
{\sl Lemma 4}.  There is a unique solution, $F(\eta,k,x)$, to the equation
$(\cL_\eta + \omega^2 ) F=0$, for which
\begin{equation}
F =(2\pi)^{-3/2}e^{ik\cdot x} +\ {\rm outgoing\ waves\/}.
\end{equation}
The map
\begin{equation}
L_2(\cM) \longrightarrow L_2({\sl \sRR}^3),
\label{l2l2}
\end{equation}
given by
\begin{equation}
f(x) \longmapsto \hat f(k) = \int dk F(\eta, k, x) f(x),
\label{hatf}
\end{equation}
is unitary.

\noindent{\sl Proof.} An immediate consequence of Lemmas 2 and 3 is
that there exists a unique outgoing smooth solution $\varphi $ to
(\ref{inhom}).
Then Eq. (\ref{fout}), relating $F$ to $F_{\rm out}$, gives us
existence and uniqueness of a smooth solution $F$ of the claimed form.
Unitarity is implied by the self-adjointness of $\cL_\eta$ for fixed
$\eta$ and the fact that $\int dk$ is a spectral measure.  A detailed
proof of unitarity, applicable with essentially trivial changes to our
case is given in Chap. 6 of Wilcox \cite{wilcox}.  (For example, the
``generalized Neumann condition,'' $\int d^3 x [f\nabla^2 g + \nabla f
\cdot \nabla g] = 0$, is replaced by $\int dV [ f e^\nu D_a (e^\nu D^a
g) +  D_a (e^\nu f) D^a (e^\nu g) ] = 0$.)
$\qquad\square$\\

\noindent{\sl Lemma 5}. For almost all $\tau$ the following holds:
Let $ a (k)\in L_{2,n} (\sRR^3)$, and let $\psi(x)=
\int dk\ F(\eta=\omega \tau, k,x) a(k)$.\\
Then $\psi(x)\in H_{n-3/2-\epsilon}(\cM_D)$, all
$\epsilon>0$.\\

\noindent{\sl Proof.}

Fourier transform $F(\eta,k,x)$, truncating smoothly at $x=R$:  With
$\chi$ the smoothed step function of Eq.~(\ref{chi}), let $g_y\in
L_2(\cM)$ be given by
\beq
g_y(x):= (2\pi)^{3/2}e^{ix\cdot y}[1-\chi(x)],
\end{equation}
with norm
\begin{equation}
\|g_y\|_{L_2(\cM)}=CR^{{3\over2}},
\end{equation}
some $C$ independent of $y$. From the fact that $F$, regarded as a map from
$L_2(\cM)$ to $L_2(\sRR^3)$ is norm-preserving, the function
\begin{equation}
\hat F(\eta,k,y):= \int dV e^{-\nu}\ F(\eta,k,x)g_y(x)
\label{hatF}\end{equation}
has norm in $k$-space
\begin{equation}
\left|\left|\hat F(\eta,\cdot,y)\right|\right|_{L_2(\sRR^3)}=CR^{3/2},
     \ \forall \eta,y.
\label{normf1}
\end{equation}
In order to bound the integral of the {\sl Lemma}, we will bound a
norm of $\hat F(\omega\tau,k,y)$ in $\tau-k$ space, for any finite
interval $I=[\tau_0,\tau_1]$ of $\tau$.  It will be convenient to take
$\tau_1= m\tau_0$, for some integer $m$.  Writing
\beq
Q(\eta, k,y)={\left|\omega^n \hat F(\eta,k,y)\right|^2 \over (1+\omega^2)^n},
\eeq
we have
\begin{eqnarray}
\int_0^{\infty} d\omega \int_I d\tau\ Q(\omega\tau,k,y)
&=& \int_0^{\infty} d\omega \int_{\omega\tau_0}^{\omega\tau_1}
d\eta \omega^{-1} Q(\eta,k,y)\nonumber\\
 &\leq& \sum_{j=0}^{\infty}\int_0^{2\pi} d\eta
\int^{2\pi(j+1)/\tau_0}_{2\pi j/
        \tau_0} d\omega\ (j+1){\tau_1\over\tau_0} \omega^{-1} Q\nonumber\\
 &\leq&  \int_0^{2\pi} d\eta \int_0^{\infty}d\omega\ \left({\tau_1\over\tau_0}
 +{\omega\tau_1\over 2\pi}\right) \omega^{-1} Q.
\label{normf2}
\end{eqnarray}
In the first inequality, we have used the fact that $Q$ is periodic in $\eta$.

{}From this relation, we obtain
\begin{eqnarray}
\int dk \int_I d\tau\ Q(\omega\tau,k,y) &=&
    \left[\int_{\omega<2\pi/\tau_0} dk
   +\int_{\omega>2\pi/\tau_0} dk\right]\int_I d\tau Q \nonumber\\
&\leq&C' + \int_{\omega>2\pi/\tau_0} dk\int_0^{2\pi} d\eta
  \left({\tau_1\over\tau_0} +
{\omega\tau_1\over 2\pi}\right) \omega^{-1} Q(\eta, k,y)
\nonumber\\
&\leq&
C' + \tau_1/\pi\int_{\omega>2\pi/\tau_0} dk\int_0^{2\pi} d\eta\ Q(\eta, k,y)
\nonumber\\
&\leq& C'+C'' R^3,
\label{normf3}
\end{eqnarray}
where the last inequality follows from Eq.~(\ref{normf1}).

Integrating over $y$, we have,
\begin{equation}
\int d\tau dk dy\
{\omega^{2n}|\hat F(\omega\tau,k,y)|^2\over(1+\omega^2)^n
 (1+y^2)^{3/2+\epsilon}}
<\infty\nonumber
\end{equation}
\begin{eqnarray}
\Longrightarrow\ & \omega^n\hat F(\omega\tau,k,y)
\in L_2(I)\otimes L_{2,-n}(\sRR^3)\otimes L_{-3/2-\epsilon}(\sRR^3)\nonumber\\
\Longrightarrow\ &\nabla^n F(\omega\tau,k,x)
\in L_2(I)\otimes L_{2,-n}(\sRR^3)\otimes H_{-3/2-\epsilon}(\cM_D)\nonumber\\
\Longrightarrow\ & F(\omega\tau,k,x)
\in L_2(I)\otimes L_{2,-n}(\sRR^3)\otimes H_{n-3/2-\epsilon}(\cM_D).\nonumber\\
\label{ineqa}
\end{eqnarray}
Thus, for almost all $\tau$,
\begin{eqnarray}
& F(\omega\tau,k,x)\in L_{2,-n}(\sRR^3)\otimes H_{n-3/2-\epsilon}(\cM_D)
\nonumber\\
\Longrightarrow\ &\int dk\ a(k) F(\omega\tau,k,x)\in H_{n-3/2-\epsilon}(\cM_D)
\label{ineqb}
\end{eqnarray}
for $a(k)\in L_{2,n}(\sRR^3)$. $\qquad\qquad\square$\\

\noindent{\sl Lemma 6}. Let $\Psi_{\rm out}$ be the outgoing field in Minkowski
space of a smooth, spatially bounded source, $\rho$.  Then data for $\Psi_{\rm
out}$ on ${\cal I}^+$ is well-defined and smooth.

\noindent {\sl Proof}.
The lemma follows from the explicit form of the flat space retarded
solution,
\begin{equation}
\Psi_{\rm out}(t=u+r,\vec x) =
\int d^3y {\rho(t=u+r-|\vec x-\vec y|, \vec y)\over{|\vec x-\vec y|}},
\label{psiouta}\end{equation}
where $r=|\vec x|$.  Writing
$|\vec x-\vec y| = r -\hat x\cdot\vec y +O(|\vec y/\vec x|),$
we have
\begin{equation}
\left| \rho(t=u+r-|\vec x-\vec y|, \vec y) -\rho(t = u+\hat x\cdot\vec y,\vec
y)
 \right| < K |\vec y/\vec x|\ {\rm max} |\dot\rho|.
\end{equation}
This bound holds for all values of $r$ along the null ray at constant
$u, \theta, \phi$, with ${\rm max} |\dot\rho|$ the maximum value of
$|\dot\rho|$ on the (compact) intersection of the support of $\rho$
with the past light cones from points of the null ray.
Then data on ${\cal I}^+$ takes the explicit form,
\begin{equation}
 \lim_{r\rightarrow\infty}r\Psi_{\rm out}(t=u+r,\vec x)
= \int d^3y \ \rho(t=u+\hat x\cdot\vec y,\vec y).
\qquad\qquad\square\label{rpsiout}\end{equation}\\

\noindent {\sl Lemma 7}.  Let $\tau$ be such that the conclusion to Lemma 5
holds.  Let
$a(k)\in L_{2,n}(\sRR^3)$, all $n\in \ZZ$, and let
\beq
\Psi (t,x) =\int dk\, a(k)F(\omega\tau ,k,x)e^{-i\omega t}. \label{Psi}
\eeq
Then $\Psi$ is smooth on $\cN$, and has data on $\cI^-$ given by Eq.\ (2.8),
\beq
\lim_{r\rightarrow\infty} r\Psi (v,r\hat r)= {i\over{(2\pi )^{1/2}}}
\int_0^\infty d\omega\, \omega a(-\omega\hat r)e^{i\omega v} . \label{data}
\eeq

That is, if $\Psi$ has for each harmonic the same ingoing part as does
a solution $\Psi_0$ in Minkowski space, then $\Psi$ has the same data
on $\cI^-$ that $\Psi_0$ has.

\noindent {\sl Proof}.
The smoothness of $a(k)$ implies, by {\sl Lemma 5}, that the integral of Eq.\
(\ref{Psi}) defines $\Psi (0,\cdot )\in H_n(\cM_D)\ \forall n$.  Since
$a(k)e^{- i\omega t}\in L_2(\sRR^3)$, we have $\Psi (t,\cdot )\in
H_n(\cM_{t,D})\ (\forall n,t)$, and $\Psi (t,\cdot )$ is smooth on each
surface $\cM_t$, because it is smooth on $\cM_{t,D}$ for all $D$.  To
see that $\Psi$ is smooth on $\cN$, note first that it is smooth on the
throat $\Sigma$ that joints $\cM_t$ to its extension $M_{t+\tau}$.
This follows from the fact that the location of the cylinder $C$
removed from $\cN$ is arbitrary:  if, instead of removing $C$ from
$\cN$, one removes a different, cylinder $C'$, disjoint from $C$,  one
obtains the same $\Psi$, because the eigenfunctions on $M$ from which
$\Psi$ is constructed are unique by Lemma 3.  Thus $\Psi (t,\cdot )$ is
smooth on each $\cM_t'$ and, in particular, on the throat $\Sigma$.

Because $\omega a(k)$ is similarly in $L_{2,n}(\sRR^3)\ \forall n$,
$\partial_t\Psi$ is smooth on each $\cM_t$ and on the throat.  Finally,
$\Psi$ is smooth on $\cN$, because $\cN$ is covered by globally
hyperbolic subspacetimes $(U, g|_U)$ which have as Cauchy surfaces
$U\cap (\cM_t \cup \Sigma\cup \cM_{t+\tau})$, for some $t$; and on $(U,
g|_U)$, $\Psi$ satisfies the smooth hyperbolic equation
$\nabla^\alpha\nabla_\alpha\Psi =0$ with smooth initial data.

We will first relate data on $\cI^+$ to $a(k)$ and then reverse the argument,
deducing the data on $\cI^-$ from that on $\cI^+$.

As in Eq.\ (\ref{fout}), we can write
\beq
F(\omega\tau ,k,x)=\chi (x)F_0(k,x)+F_{\rm out}(\omega\tau ,k,x) ,
\eeq
where $F_0(k,x)=(2\pi )^{-3/2}e^{ik\cdot x}$.  Then
\beq
\Psi = \chi (x)\Psi_0^- +\Psi_{\rm out},
\eeq
where
\beq
\Psi_{\rm out} = \int dk\, a(k)F_{\rm out}(\omega \tau ,k,x)e^{-i\omega t} ,
\quad
\Psi_0^- = \int dk\, a(k)F_0(k,x)e^{-i\omega t}
\label{psiout}\eeq
The integral defining $\Psi_{\rm out}$ converges in $L_2(\cM_D)$ to a
smooth function, because the integrals defining $\Psi$ and the
Minkowski-space solution $\Psi_0$ so converge.  $\Psi$ and $\Psi_{\rm
out}$ are smooth on $\cM$ because they are smooth on $\cM_D$ for all
$D$.

We can use the spherical harmonic basis $\{ Y_{lm}\}$ for $L_2(S^2)$ to write,
for the exterior region ${\cal E}$,
\begin{eqnarray}
F_0(k,x) &=& \left({2\over\pi}\right)^{1/2}
\sum_{lm} i^l j_l(\omega r) Y^*_{lm}(\hat k) Y_{lm}(\hat x)
\\
F_{\rm out}(\eta ,k,x) &=& \sum_{lm} \gamma_{lm} (\eta ,k)h_l^{(1)}(\omega
r)Y_{lm}(\hat x).
\end{eqnarray}
with the sums converging in $L_2(S^2)$.  Then
\beq
\Psi_{\rm out} = {1\over{\sqrt{2\pi}}} \int d\omega\, \omega^2 \sum_{lm}
c_{lm}(\omega )h_l^{(1)}(\omega r)Y_{lm}(\hat x)e^{-i\omega t},
\eeq
where
\beq
c_{lm}(\omega ) = \sqrt{2\pi}\int d\Omega_k\, a(k)
         \gamma_{lm}(\omega\tau, k).
\eeq
We can similarly rewrite the convergent integrals for $\Psi_0$ and $\Psi$ in
${\cal E}$:
\beq
\Psi_0^- = {1\over{\sqrt{2\pi}}} \int d\omega\, \omega^2
\sum_{lm} 2a_{lm}(\omega )j_l(\omega r)Y_{lm}(\hat x)e^{-i\omega t},
\label{Psio} \eeq
\beq
\Psi = {1\over{\sqrt{2\pi}}} \int d\omega\,
\omega^2\sum_{lm}[b_{lm}(\omega )h_l^{(1)}(\omega r)+a_{lm}(\omega
)h_l^{(2)}(\omega r)]Y_{lm}(\hat x)e^{-i\omega t},
\eeq
where
\beq
a_{lm}(\omega) = i^l\int d\Omega_k\, a(k) Y^*_{lm}(\hat k) \label{alm}
\eeq
and
\beq
b_{lm}=a_{lm}+c_{lm}. \label{blm}
\eeq

The construction of $\Psi_{\rm out}$ from outgoing waves $F_{\rm out}$
of Eq. (\ref{fout}), satisfying the inhomogeneous equation
(\ref{inhom}) expresses  $\Psi_{\rm out}$ on $\cE$ as the retarded
solution on Minkowski space, (\ref{rpsiout}), to $\square \Psi_{\rm
out}=\rho$, with $\rho:=\square\Psi_0 \chi$.  If one writes $\rho$ of
Eq.\ (\ref{rpsiout}) as a sum of spherical harmonics and uses
Eq.\ (\ref{psiouta}) to relate $c_{lm}$ to $\rho_{lm}$, data
(\ref{rpsiout}) on $\cI^+$ for $\Psi_{\rm out}$ becomes
\beq
\lim_{r\rightarrow\infty} r\Psi_{\rm out} (t=u+r,\hat x)
= {1\over{\sqrt{2\pi}}} \int
d\omega\, \omega \sum_{lm} i^{-(l+1)} c_{lm}(\omega )Y_{lm}(\hat x)e^{-i\omega
u}.
\eeq
A finite-energy solution on Minkowski space of the form (\ref{Psio})
has data on $\cI^+$ given by
\beq
\lim_{r\rightarrow\infty} r\Psi_0^-(u+r,\hat x)={1\over{\sqrt{2\pi}}}
\int d\omega\, \omega \sum_{lm} i^{-(l+1)}a_{lm}(\omega )
Y_{lm}(\hat x)e^{-i\omega u}.
\eeq

The first part of the proof of Lemma 3 implies for the solution to
$\cL_\eta\Psi =0$ on ${\cal E}$, eigenfunctions of $\cL_\eta$ satisfy,
for each $\omega$,
\beq
\sum_{lm} |a_{lm}(\omega )|^2 = \sum_{lm} |b_{lm}(\omega )|^2 .
\eeq
Thus the data induced by $\Psi$ on $\cI^+$ satisfies the constraints
imposed on $a(k)$:  $b(k)\in L_{2,n}(\sRR^3 )$ all $n\in\ZZ$.  As a
result, the argument just given, with $\cI^-$ and $\cI^+$ reversed,
implies
\beq
\Psi = \Psi_0^++\Psi_{in} ,
\eeq
with $\Psi_0^+$ and $\Psi_{in}$ smooth, and given on ${\cal E}$ by expressions
\begin{eqnarray}
\Psi_0^+ &=& {1\over{\sqrt{2\pi}}} \int d\omega\, \omega^2 \sum_{lm}
2b_{lm}(\omega )j_l(\omega r)Y_{lm}(\hat x)e^{-i\omega t},\\
\Psi_{in} &=& {-1\over{\sqrt{2\pi}}} \int d\omega\, \omega^2 c_{lm}(\omega
)h_l^{(2)}(\omega r)Y_{lm}(\hat x)e^{-i\omega t} ,\\
\end{eqnarray}
converging in $L_2(\cM_D)$ for each $D$; and data on $\cI^-$ is well-defined,
with
\beq
\lim_{r\rightarrow\infty} r\Psi (v-r, r\hat x)={1\over{\sqrt{2\pi}}} \int
d\omega\, \omega \sum_{lm} i^{(l+1)}
(b_{lm}+c_{lm})Y_{lm}(\hat x)e^{-i\omega v},
\eeq
whence, using Eq.\ (\ref{blm}), we recover (\ref{data}).

\noindent
{\sl Lemma 8.} Let $\psi$ be a smooth retarded solution on Minkowski space
to the massless scalar wave equation whose source $\rho$ has support
within a compact spatial region $r<R$, and suppose that $\psi$ has zero
data on $\cI^-$ and data $f$ on $\cI^+$ with finite energy norm,
$||f||_{H_1(\cI^+)}$. Then $\psi$ has finite energy norm
$||\psi||_{H_1(\cH)}$ on a spacelike hyperplane $\cH$ of Minkowski
space.\\

\noindent
{\sl Proof}. The proof will follow from conservation of energy, but,
because the energy density of a massless scalar field does not include
a term proportional to $\psi^2$, we will need to consider both the
energy of $\psi$ and of a time integral of $\psi$ to bound $||\psi||_1$
on a spacelike hyperplane.

Let $\cH$ be the surface $t=0$.  We need only consider the retarded
field of a source $\rho$ that vanishes for $t>0$, because the field on
$t=0$ depends only on values of $\rho$ for $t<0$. Let $t^\alpha$ be the
timelike Killing vector $\partial_t$, and let $J^\alpha =
-T^\alpha_\beta t^\beta$, with $T^\alpha_\beta$ the energy-momentum
tensor of $\psi$.  Let $E(u_0)$ be the energy of $\psi$ on the part
$\cH(u_0)$ of $\cH$ with $r<|u_0|$:
\beq
E(u_0) = \int_{\cH(u_0)} dS_\alpha J^\alpha,
\eeq
with $dS_\alpha$ along the normal $\nabla_\alpha t$
Let $\cI_{u_0,v_0}$ be the part of the past null cone $v=v_0$ lying to the
future of $t = (u_0+v_0)/2$; and let $\cJ_{u_0,v_0}$ be the part of the future
null cone $u=u_0$ between $t=0$ and $t = (u_0+v_0)/2$.  Then $\cH(u_0)$,
$\cI_{u_0,v_0}$, and $\cJ_{u_0,v_0}$ form the boundary of a source-free region
of Minkowski space.  Denote by $F_I(u_0,v_0)$ and $F_J(u_0,v_0)$ the flux
through $\cI_{u_0,v_0}$ and $\cJ_{u_0,v_0}$, respectively,
\beq
F_I=\int_{\cI_{u_0,v_0}}
dS_\alpha J^\alpha, \qquad F_J = \int_{\cJ_{u_0,v_0}} dS_\alpha J^\alpha,
\label{fifj}\eeq
choosing $dS_\alpha$ along the normals $\nabla_\alpha u$ and
$\nabla_\alpha v$. Then $\nabla_\alpha J^\alpha = 0$ implies
\beq
E(u_0)+F_J(u_0,v_0) = F_I(u_0,v_0).
\label{efj}\eeq

A massless scalar field, $\psi$ satisfies the
dominant energy condition:  the vector $J^\alpha$ is future-directed
non-spacelike.  Consequently $E(u_0),\ F_I(u_0,v_0)$, and $F_J(u_0,v_0)$ are
all
positive, and we have
\beq
E(u_0)\leq F_I(u_0,v_0).
\label{eu}\eeq
The retarded field of a smooth,
spatially compact source has asymptotic behavior
\begin{eqnarray}
\psi(u,x) &=& f(u,\hat x)/r + O(r^{-2}),\nonumber\\
\nabla^\alpha v\nabla_\alpha \psi &=& \partial_u f(u,\hat x)/r
         + O(r^{-2}),\nonumber\\
h^{\alpha\beta}\partial_\beta \psi &=& O(r^{-2}),
\label{asymp}\end{eqnarray}
where $h^{\alpha\beta}=\eta^{\alpha\beta}+t^\alpha t^\beta$ is the
projection orthogonal to $t^\alpha$. Consequently,
\beq
\lim_{v_0\rightarrow\infty}F_I(u_0,v_0) = F_{\cI^+(u_0)},
\label{limfi}\eeq
where $F_{\cI^+(u_0)}$ is the flux through the part of ${\cI^+(u_0)}$ lying
to the future of $u=u_0$:
\beq
F_{\cI^+(U)}=\int_{-u_0}^0 du \int d\Omega\ |\partial_u f(u,\hat x)|^2.
\label{fscri}\eeq
Since $E(u_0)$ is independent of $v_0$, we have for all $u_0$,
$E(u_0)<F_{\cI^+(u_0)}$.
Hence the scalar field $\psi$ has finite energy on $\cH$:
\beq
E = \lim_{u_0\rightarrow -\infty} E(u_0) \leq
\lim_{u_0\rightarrow -\infty} F_{\cI^+(u_0)}<||f||_{H_1(\cI^+)}.
\label{e}\eeq

In order to bound $\int_\cH dV\ \psi^2$, we essentially repeat the argument for
\beq
\tilde\psi = \int_0^u du'\ \psi(u',x).
\eeq
Eq.~(\ref{asymp}) implies that $\tilde\psi$ has data $\tilde f$ on $\cI^+$,
where
\beq
\tilde f = \int_0^u du'\ f(u,\hat x).
\eeq
Bounding the energy $\tilde E$ of $\psi$ on $\cH$ bounds the $L_2$ norm of
$\psi$, because
\begin{eqnarray}
\tilde E &=& \int_\cH dS_\alpha \tilde J^\alpha = \int_\cH dV
{1\over2}[(\partial_u\tilde\psi)^2 +
(\partial_u\tilde\psi-\partial_r\tilde\psi)^2 +
r^{-2}(\partial_\theta\tilde\psi)^2+(r
\sin\theta)^{-2}(\partial_\varphi\tilde\psi)^2] \nonumber\\
&\geq& \int_\cH dV {1\over2} \psi^2.
\label{etilde}\end{eqnarray}

Because $\psi$ vanishes for $u\geq 0$,  $\tilde\psi$ satisfies the scalar wave
equation with source $\tilde\rho$:
Using
\beq
\square = -2\partial_u\partial_r +\partial_r^2 +{1\over
r^2}(\partial_\theta^2+{1\over \sin^2\theta}\partial_\varphi^2),
\eeq
and
\beq
\partial_u \tilde\psi(u,x) = \psi(u,x) = \int_0^u du'\ \partial_{u'}
\psi(u',x),
\eeq
we have
\begin{eqnarray}
\square \tilde\psi &=& \int_0^u du'\ \square\psi(u',x)
\nonumber\\
&=&\tilde \rho(u,x).
\end{eqnarray}
Then Equations (\ref{fifj}, \ref{efj}, \ref{eu}) hold for the energy fluxes
$\tilde F_I(u_0,v_0)$, and $\tilde F_J(u_0,v_0)$, and, in particular,
\beq
\tilde E(u_0)\leq \tilde F_I(u_0,v_0).
\label{eutilde}\eeq

However, $\tilde\psi$ is not the {\it retarded} solution to the scalar
wave equation with source $\tilde\rho$, so we must check that the
asymptotic conditions (\ref{asymp}) on $\psi$ hold for $\tilde\psi$.
This is easy:  A function $O(r^{-n})$ has, for $r$ greater than some
$r_0$,  the form $g(u,x)/r^n$ where, for fixed $u, \hat r$, $g$ is a
bounded function of $r$.  In our case, $\psi$ is smooth, and the
corresponding functions $g$ are smooth and bounded in a compact domain
$[0,u]$ for $u$ (and hence for $u, \hat r$).  Thus $\int_0^u g(u')du' =
O(r^{-n})$, and it follows that $\tilde\psi$ satisfies (\ref{asymp}).
The analogues of Eqs.~(\ref{fscri}) and (\ref{e}),
\begin{eqnarray}
\tilde F_{\cI^+(u_0)}&=&\int_{u_0}^0 du \int d\Omega |\partial_u \tilde
f(u,\hat x)|^2\nonumber\\
&=&\int_{u_0}^0 du \int d\Omega |f(u,\hat x)|^2 < \infty,
\end{eqnarray}
and
\beq
\tilde E = \lim_{u_0\rightarrow -\infty} \tilde E(u_0) \leq
\lim_{u_0\rightarrow-\infty} \tilde F_{\cI^+(u_0)}<||f||_{L_2(\cI^+)},
\eeq
together with Eq.~(\ref{etilde}), imply $||\psi||_{L_2(\cH)}<\infty$.
Finally, Eq.~(\ref{e}) and the bound on $||\psi||_{L_2(\cH)}$ implies
$||\psi||_{H_1(\cH)}<\infty$.  $\qquad\qquad\square$

\noindent
{\sl Corollary} Under the assumtions of {\sl Lemma 7}, $\Psi_{\rm out}$ of
Eq.~(\ref{psiout}) is asymptotically regular at spatial infinity.  \\

\noindent
{\sl Proof}.  As in {\sl Lemma 7}, $\Psi_{\rm out}$ is on $\cE$ the
retarded solution of Eq. (\ref{psiouta}) with smooth spatially bounded
source $\rho$. Then {\sl Lemma 8} implies $\Psi_{\rm out}$ is regular
at spatial infinity. \\

\noindent
{\sl Proof of Proposition 1}.  The proof is essentially immediate from
the Lemmas.  By {\sl Lemmas 6} and {\sl 7}, with $\Phi=2 Re\Psi$, for
almost all $\tau$ $\Phi$ is a smooth solution on $\cN$ to the scalar
wave equation, with data $f$ on $\cI^-$.  Finally, on $\cE$, $\Psi =
\Psi_0 +\Psi_{\rm out}$ where $\Psi_0$ is asymptotically regular at
spatial infinity because it is a solution to the flat-space wave
equation with data having finite energy-norm on $\cI^-$ and $\Psi_{\rm
out}$ is regular by the {\sl Corollary} to {\sl Lemma 8}.
$\qquad\qquad\square$

\subsection{Restricted uniqueness theorem for a massless scalar field}
\label{sec:unique}

Because the system is linear, uniqueness means that the only solution
to Eq.~(\ref{kg}) with zero data at $\cI^-$ is $\Phi = 0$.  This is not
true in the geometrical optics limit, because closed null geodesics
$c(\lambda)$ can loop through the wormhole and never reach $\cI$; and
one might worry that there are analogous smooth solutions that vanish
in past and future and have zero data on $\cI^-$.  The following
restricted uniqueness theorem rules them out:  smoothed versions of a
looping zero-rest mass particle spread and reach $\cI$.  Denote by
${\cal E}_{K_t}$  an energy norm of the field on a compact submanifold
$K_t\subset\cM_t$, where manifolds in the family $\{K_t\}$ are related
by time-translation along the Killing trajectories:
\begin{equation}
{\cal E}_{K_t} = \int_{K_t} dV e^{-\nu} \ [\ |\nabla\Phi|^2 + |\Phi|^2] =
\left|\left|\Phi\right|\right|_{H_1 (K_t)}.
\end{equation}
\vspace{0.2in}

\noindent
{\sl Proposition 2}.  If $\Phi(t,x)$ is a smooth solution
to $\nabla_\a\nabla^\a\Phi=0$, having finite energy and zero initial data on
$\cI^-$, and if $\lim_{t\rightarrow\pm\infty}{\cal E}_{K_t} = 0$ for any family
of
compact (time-translation related) $K_t$, then $\Phi = 0$.\\

\noindent
{\sl Proof}.  The result is a corollary of Rellich's uniqueness theorem
for each mode.  The requirement that the energy norm on any compact $K$
vanishes as $t\rightarrow\infty$ allows one to transform the solution as
follows.  Denote by $H_T$ a smoothed step function,
\begin{eqnarray}
H_T(t) & = &\left\{
\begin{array}{l}
	      1,\;|t| \leq T\\
              0,\;|t| \geq T+\epsilon,
\end{array}\right.
\label{step2}\end{eqnarray}

and write the Fourier transform of $\Phi$ in the form
\begin{equation}
\hat\Phi(\omega,x) = \lim_{T\rightarrow\infty}
                   (2\pi)^{-1/2}\int dt H_T(t) e^{-i\omega t} \Phi(t, x).
\label{hatphi}\end{equation}
We have
\begin{equation}
0=\int dt H_T(t) e^{-i\omega t} \nabla_\a\nabla^\a\Phi(t,x)
=\int e^{-\nu}\ dt\ e^{-i\omega t}[H_T (\omega^2+ \cL)
+ 2i\omega\partial_t H_T - \partial^2_t H_T]\Phi(t),
\label{ht}
\end{equation}
whence
\begin{equation}
\left|\int dt e^{-i\omega t}[H_T (\omega^2+ \cL)\Phi(t)\right|^2
\leq \left[ \int\ dt\ (| 2\omega\partial_t H_T\Phi(t)| +|\partial^2_t
H_T\Phi(t)| ) \right]^2.
\label{htw}
\end{equation}
Because $ \partial_t H_T$  and $\partial^2_t H_T$ are bounded functions
of $t$ with compact support, we can write
\begin{equation}
\int_{K} dV \left|\int dt H_T e^{-i\omega t}(\omega^2+ \cL)
\Phi(t,x)\right|^2\leq C \max_{|t|\in[T,T+\epsilon]}\int_{K} dV|\Phi(t,x)|^2,
\end{equation}
or
\begin{equation}
\left|\left|\int dt H_T e^{-i\omega t}(\omega^2+ \cL)\Phi(t,x)
\right|\right|^2_{L_2(K_t)}\leq C
\max_{t\in[T,T+\epsilon]}\|\Phi\|^2_{L_2(K_t)},
\end{equation}
where $C$ is a constant independent of $T$.  From our assumption that
$\left|\left|\Phi\right|\right|^2_{L_2(K_t)} \rightarrow 0$ as
$t\rightarrow\infty$, we have
\begin{equation}
\lim_{T\rightarrow\infty}\max_{t\in[T,T+\epsilon]}\left|\left|\Phi\right|\right|^2_{L_2(K_t)}=0
\end{equation}
implying
\begin{equation}
\lim_{T\rightarrow\infty}\left|\left|\int dt H_T e^{-i\omega t}
      (\omega^2+ \cL)\Phi(t,x)\right|\right|^2_{L_2(K_t)}=0.
\end{equation}
It follows that for all $\omega$, $\int dt e^{-i\omega t}(\omega^2+
\cL)\Phi(t,x)=(\omega^2+
\cL)\hat\Phi(\omega,x)$ vanishes for almost all $x$.
Finally, since finite energy and zero initial data imply purely outgoing
at spatial infinity, by Rellich's theorem $\Phi$ must vanish as well.
$\qquad\qquad\square$

\subsection{Other massless fields}
\label{sec:other}

   Extending these results to Weyl and Maxwell fields appears
straightforward, at least for hyperstatic spacetimes, with $t^\a t_\a =
-1$.  The statement of the existence theorem is identical, with the
scalar field $\phi$ replaced by a Weyl spinor $\phi^A$ and (say) a
vector potential $\phi^\alpha$ for a free Maxwell field
$F^{\alpha\beta}$ in a Lorentz gauge with $A_\a t^\a = 0$.  The
statement of uniqueness for a Weyl field is identical to that for a
scalar field; for a Maxwell field it must be modified to exclude the
time independent solutions that have nonzero flux threading the
handle:\\

If $F^\alpha\beta$ is a smooth solution to
\begin{equation}
\nabla_\alpha F^{\alpha\beta}=0,\qquad \nabla_{[\alpha} F_{\beta\gamma]}=0
\end{equation}
with
$$\int_S F_{\alpha\beta}dS^{\alpha\beta}=0
 =\int_S\ ^*F_{\alpha\beta}dS^{\alpha\beta},$$
having finite energy and zero initial data on $\cI^-$, and if
$\lim_{t\rightarrow\pm\infty}{\cal E}_{K_t} = 0$ for any family of
compact (time-translation related) $K_t$, then $F^{\alpha\beta} = 0$.

   For a hyperstatic spacetime, the proofs appear to require only minor
changes, because we can again decompose the wave operator for vector
and spinor fields in the manner
\beq
\nabla_\beta \nabla^\beta  = -\mbox{\pounds}_t^2 + \cL,
\label{defl3}\eeq
with $\cL = D^a D_a$, as in Eq.~\ref{defl2}. Then the harmonic
components $\phi^\a,\ \phi^A$ satisfy
\beq
(\omega^2 + \cL)\phi^\a = 0, \qquad (\omega^2 + \cL)\phi^A = 0.
\label{wave}\eeq
The operator $\cL$ is symmetric on $L_2(\cM)$, defined for vector and
spinor fields $\phi$ on $\cM$ by the norm $ <\phi|\phi> = \int_\cM dV e^{-\nu}
|\phi|^2$, with $|\phi|^2\equiv \bar\phi^\a \phi_a$ and $|\phi|^2\equiv
t_{AA'}\bar\phi^{A'} \phi_A$ for vectors and spinors, respectively.  The
definition of the Sobolev spaces of Sect.~\ref{sec:sobolev} are similarly
extended automatically to vectors and spinors by contracting vector
indices with $h_{ab}$ and spinor indices with $t_{AA'}$ (In a spinor
frame associated with an orthonormal frame for which $t^\a$ is the
timelike frame vector, $t_{AA'}$ has components $t_{II'} = \delta_{II'}$.
Sobolev spaces of Sect.~\ref{sec:sobolev} are similarly extended
to vector and spinor fields.  Finally, the space $\HH_2$, is defined
for vectors and spinors as the set of fields in $H_2(\cM)$ satisfying
the boundary conditions (\ref{vbc},\ref{wbc}).

Here is a sketch of the proofs.   The existence theorem involves the
same set of lemmas.\\
{\sl Lemma 1}, self-adjointness of $\cL_\eta$ on the spaces $\HH_2$ of
fields in $H_2(\cM_t)$ again follows from the symmetry of $\cL_\eta$,
reflecting the fact that boundary conditions imply smoothness of the
fields $\hat \phi^A$ and $\hat \phi^\alpha$ on $\hat M$. \\
{\sl Lemma 2} and its proof can be repeated as written with $\phi$ and
$\rho$ regarded as vectors or spinors.\\
In the proof of {\sl Lemma 3}, each of the components of $\phi^\a$
($\phi^A$) with respect to covariantly constant frames (spinor frames)
on the exterior region ${\cal E}$ are scalars satisfying
Eqs.~(\ref{hlm1}), (\ref{hlm2}), and (\ref{hlm3}).  They therefore
vanish on ${\cal E}$ and, by elliptic continuation, vanish on all of
$\cM$.  \\
In {\sl Lemma 4}, eigenfunctions for vector and spinor fields are defined by
\begin{mathletters}\label{fk2}\begin{eqnarray}
F^A &=\ \chi (2\pi)^{3/2}\epsilon^A e^{ik\cdot x} +\varphi^A \label{varphib} \\
F^\alpha &=\ \chi(2\pi)^{3/2}\epsilon^\alpha e^{ik\cdot x}+\varphi^\alpha,
\label{varphic}
\end{eqnarray}
\end{mathletters}
where $\epsilon^\a$ and $\epsilon^A$ are covariantly constant on ${\cal E}$ and
satisfy $k_\a \epsilon^\a = 0 = k_{AA'}\epsilon^A$.
The first part of {\sl Lemma 4}, existence and uniqueness of
eigenfunctions $F^\a, F^A$, is again immediate from {\sl Lemmas 2} and
{\sl 3}.  The second part, unitarity, again appears to be a
straightforward extension of Chap. 6 in Wilcox \cite{wilcox}, but here
there are details we have not checked.  \\
The proof of {\sl Lemma 5}
goes through as written if $a$ in {\sl Lemma 5} is
interpreted as a vector (spinor) with a covariant index and their
product is read as a dot product: e.g., $f(x)=\int dk\ F^\a(\eta=\omega
\tau , k,x) a_\a(k)$\\
In {\sl Lemma 6}, the equations are correct as written for the components
of $\phi^\a$ and $\phi^A$, but one must use, in addition, the fact that the
fields satisfy the peeling theorem for flat space \cite{penrose} to
complete the characterization of their behavior.
In {\sl Lemma 7}, the proof of smoothness can be read as written.  The
proof of regularity at $\cI^-$ and the recovery of initial data relies
on a spherical harmonic decomposition that can be modified in a
standard way for spinors and vectors.  Finally, {\sl Lemma 8} and the proof
of Propositions 1 and 2, can be read as written, with the change in the
statement of Proposition 2 given above.

\section{The Cauchy problem for more general spacetimes}
\label{sec:general}

The work reported above shows the existence of an unexpected class of
spacetimes for which an existence theorem and at least
a partial uniqueness theorem can be proved.
How broad is the class of spacetimes for which a
generalized Cauchy problem is well-defined?
Examples of spacetimes with
CTCs for which one can prove existence and uniqueness for linear wave
equations are not difficult to find, if one allows singularities and
does not require that solutions have finite energy \cite{politzer}, and
we will display some examples below.  Finding examples of nonsingular
geometries with CTCs and a well-defined Cauchy problem is more
difficult, but the earlier work by Morris {\it et.  al.} \cite{mty} is
persuasive:  Their time-tunnel examples are asymptotically flat
spacetimes in which CTCs are confined to a compact region and for which
there appears to be a well-defined initial value problem for data on a
spacelike hypersurface to the past of the {\it nonchronal region}, the
set of points through which CTCs pass.  In the present section we
present a uniqueness result complementary to that of
Sect.~\ref{sec:unique}, a conjecture on existence and uniqueness, and
examples of spacetimes that do or do not have a well-posed initial
value problem.

\subsection{A result on uniqueness for spacetimes with a compact
nonchronal region}
\label{sec:unique2}

In Sect.~\ref{sec:unique} we ruled out, for the static spacetimes
considered, a lack of uniqueness corresponding to a field forever
trapped {\it inside} the nonchronal region, a smooth analog of a closed
null geodesic.  Here we show, for spacetimes with a compact nonchronal
region, $\cA$, that when solutions to the Cauchy problem exist, they
are unique {\it outside} $\cA$.  Because data is now given on a
spacelike surface (instead of $\cI^-$), we need no longer restrict
consideration to massless wave equations.  {\sl Initial data} on a
spacelike hypersurface $\cM$ for a solution $\Phi$ to the scalar wave
equation,
\beq
(-\nabla_\a \nabla^\a + m^2)\Phi = 0,
\label{massive}\eeq
will mean the pair of functions
\beq
\phi = \Phi|_\cM,\ \pi=n^\a\nabla_a\Phi|_\cM ,
\label{piphi}\eeq
where $n^\a$ is a unit normal to $\cM$.\\

\noindent
{\sl Proposition 3}.  Let $\cN, g_{\alpha \beta}$ be a smooth, asymptotically
flat spacetime with regions $\cF$ and $\cP$ to the future and past of a compact
4-dimensional submanifold $\cA$ defined by $\cF = \cN \backslash J^- (\cA) $
and $\cP = \cN \backslash J^+ (\cA) $, where both $\cF$ and $\cP$ are globally
hyperbolic and foliated by complete spacelike 3-manifolds.
Suppose for arbitrary smooth data with finite energy and a Cauchy surface
$ \cM_{P}$ of $\cP$ that the scalar wave equation has a solution
on $\cN$ with finite energy on $\cF$, and suppose that for arbitrary
smooth data with finite energy and a Cauchy surface $\cM_F$ of $\cF$
that the scalar wave equation has a solution on $\cN$ with finite energy on
$\cP$.  Then, solutions on $\cN$ with finite energy in $ \cN \backslash \cA $
are unique in $\cN \backslash \cA $.

\noindent
{\sl Proof}.  The proof relies on the nondegeneracy of the symplectic
form,
\beq
\omega_\cM (f,g):=\int_\cM dS_\a (f\ \nabla^\a g - g\ \nabla^\a f),
\label{omega1}
\eeq
and the fact that $\omega$ is independent of hypersurface.  That is,
let $\cB\subset\cN$ be a slab, a 4-dimensional submanifold of $\cN$
bounded by two submanifolds $\cM$ and $\cM'$ in the foliations of
$\cP$ and $\cF$ that coincide outside of a compact region.  Then, for any
two solutions with finite energy,
\begin{eqnarray}
0 = &\int_{\cB} d^4V f(\nabla_\a \nabla^\a - m^2)g \cr
  = &\int_{\partial\cB} dS_\a (f\ \nabla^\a g - g\ \nabla^\a f)\cr
  = &\omega_\cM (f,g) - \omega_{\cM'} (f,g).
\label{omega2}
\end{eqnarray}

Suppose the theorem is false.  Then there is a solution
$\Phi$ to Eq.~(\ref{massive}) with zero initial data on
$\cM_P$ (say) and with $\Phi$ nonzero and with finite energy somewhere on
$\cN\backslash\cA$.
Thus a hypersurface $\cM$ in the foliation $\cF$ has nonzero
data for $\Phi$ and we can deform $\cM$ outside the support of
$\Phi$ (in the intersection of $\cF$ and $\cP$) to coincide with $\cM_P$.
Because $\omega$ is non-degenerate,
there is data $(\Psi, \dot\Psi)$ on $\cM$ such that
\beq
\omega_\cM (\Phi,\Psi) \neq 0.
\eeq
But, by hypothesis, a solution $\Psi$ exists on $\cN$, corresponding to
the initial data on $\cM$; and the fact that $\omega$ is independent of
hypersurface implies $\omega_{\cM_P}(\Phi,\Psi) \neq 0$,
contradicting the assumption that $\Phi$ has vanishing initial data on
$\cM_P$.   $\qquad\qquad\square$\\

\noindent
{\sl Corollary}.  {\sl Proposition 3} holds for the Maxwell,
Dirac, and Weyl fields.

\noindent
{\sl Proof.} Each of the three fields has a conserved symplectic
product $\omega$.  The proof goes through as stated, with the
symplectic product and initial data of a scalar field replaced by that
of the Maxwell and Dirac fields and with the energy norm for a
scalar field replaced by vector and spinor energy norms.
$\qquad\qquad\square$

Because the billiard-ball examples considered by Echeverria {\sl et.
al.}\cite{fmetal,ekt91} have a multiplicity of solutions for
the same initial data, uniqueness in spacetimes with CTCs is likely to
hold only for free or weakly interacting fields.  Because solutions
seem always to exist for the billiard ball examples in the spacetimes
they considered, it may be that classical interacting fields
have solutions on spacetimes for which solutions to the
free field equations exist.

\subsection{A conjecture}
\label{sec:conjecture}

The solution to the problem posed at the beginning of this section -- to
delineate the class of spacetimes for which a generalized Cauchy
problem is well-defined -- is, of course, not known.  We present a
conjecture here, motivated by examples of geometries which appear to
have a well-defined Cauchy problem and by examples of others for which either
existence or uniqueness fails.  We will motivate the conjecture with a
brief reminder of some examples that are by now well known; a more
detailed discussion of additional spacetimes will be given in
Sect.~\ref{sec:examples}.

A helpful 2-dimensional example of a spacetime where the Cauchy problem
is not well defined is Misner space.  This is the quotient of the half
of Minkowski space on one side of a null line $L$ by the subgroup $
\{1, B^{\pm 1}, B^{\pm 2}, \dots\}$ generated by a boost $\cB $ about a
point of the null line.  Equivalently, if $\ell$ is a null line
parallel to $L$, and $\cB \ell$ is its boosted image, then Misner space is
the strip betwee $\ell$ and $\cB \ell$, where boundary points related by $\cB $
are identified (see Fig.~\ref{misner}).  Misner space has a single
closed null geodesic, ${\cal C} = CC'$ in the Figure, and the past
$\cP$ of ${\cal C}$ is globally hyperbolic. The future of ${\cal C}$ is
nonchronal, so ${\cal C}$ is a chronology horizon, a Cauchy horizon
that bounds the nonchronal region.  Initial data for the scalar-wave
equation can be posed on a
Cauchy surface $\cM$ of $\cP$, but solutions have divergent energy on the
chronology
horizon.

The reason solutions diverge is clear in the geometrical optics limit.
A light ray $\gamma$, starting from $\cM$, loops about the space and
is boosted each time it loops.  Because $\gamma$ loops an infinite
number of times before reaching ${\cal C}$, its frequency and energy
diverge as it approaches the horizon.  The ray $\gamma$ is an
incomplete geodesic: it reaches the horizon in finite affine parameter
length, because each boost decreases the affine parameter by the
blueshift factor $\a\equiv [(1+V)/(1-V)]^{1/2}$, with the velocity of
the boost $V$.\footnote
{That is, trajectories of a (locally-defined)
timelike Killing vector cross the null geodesic at a sequence of
points. The Killing vector can be used to compare the affine parameter
at succesive crossing points by time-translating a segment of the
geodesic to successively later segments.  Compared in this way, the
affine parameter of a given segment will will be less than that of the
next segment by the blueshift factor $[(1+V)/(1-V)]^{1/2}$.}
This behavior is not unique to Misner space: A theorem due to Tipler
\cite{tipler} shows that geodesic incompleteness is generic in
spacetimes like Misner space in which CTCs are ``created'' --
spacetimes with a nonchronal region to the future of a spacelike
hypersurface.

In more than two dimensions, however, the existence of
incomplete null geodesics like $\gamma$ does not always imply that the
chronology horizon is unstable.  This is because there may be
only a set of measure zero of such geodesics so that the energy
may remain finite on the chronology horizon.  For the time-tunnel spacetimes
considered in refs \cite{mty,fmetal}, a congruence of null rays
initially parallel to $\gamma$ spreads as the rays approach the
chronology horizon. When the spreading of the rays overcomes the
successive boosts (when the fractional decrease in flux is greater than
the fractional increase in squared frequency), the horizon is stable in
the geometrical optics approximation.  Because the instability of the
chronology horizon (or of the spacetime to its future) appears to be
the obstacle to a well defined Cauchy problem, we are led to the
following conjecture.

\noindent
{\sl Conjecture}.  Consider a spacetime $\cN, g_{\alpha \beta}$ that is \\
(i) a smooth, asymptotically flat, and for which past and future regions
$\cP = \cN \backslash J^+ ( \cA)$ and $\cF = \cN \backslash J^- (\cA) $ of
a compact 4-dimensional submanifold $\cA$ are
globally hyperbolic. Suppose that \\
(ii) the Cauchy problem for massless fields is well-defined in the
geometric optics limit.
\\
Then the Cauchy problem for massless wave equations (for scalar,
Maxwell, and Weyl fields) is well-defined.

Because the instability of massive fields also corresponds in the geometric
optics limit to an instablility of particles moving along trajectories
that become null as one approaches the chronology horizon, it may be
that massive wave equations also have a well-defined Cauchy problem for
the same spacetimes.

\subsection{Examples and counterexamples}
\label{sec:examples}

For no asymptotically flat spacetime in 4-dimensions, in which CTCs are
confined to a compact region, are we aware of a rigorous demonstration
that finite-energy solutions to the scalar wave equation do exist for
arbitrary initial data, or that solutions are unique.  There may also
be no published counterexamples, 4-dimensional asymptotically flat
spacetimes with a compact nonchronal region (more precisely, spacetimes
satisfying condition (i) of the conjecture) for which the nonexistence
or nonuniqueness of solutions to free-field equations is proven; but
counterexamples like this are not difficult to find.  We present below
two examples of asymptotically flat spacetimes which are globally
hyperbolic to the past and future of a compact region; one can prove
for one of them that no solution exists for generic data and for the
other that solutions are not unique.  The examples show that, without
some requirement akin to the well-defined geometric optics limit of our
conjecture,
there can be too many closed geodesics to allow a well-defined initial
value problem.  It might also be straightforward to show, for the
time-tunnel spacetimes of Refs.~\cite{mt,mty,fmetal}, that whenever the
chronology horizon is unstable in the geometric optics limit, it is genuinely
unstable
for fields with smooth initial data.

The first example is a time-tunnel spacetime like those of Morris {\it
et. al.} but with a metric that is everywhere smooth and is chosen to
induce a flat 2-metric on the part of the the identified spheres that
face each other.  The spacetime is also flat between the flat pieces of
the spheres, so the geometry includes a region isometric to a piece of
(Misner space)$\times E^2$, where $E^2$ is flat Euclidean 2-space.
For definiteness, we shall identify it with the following piece of
Minkowski space.  Regard Misner space as the strip of 2-dimensional
Minkowski space between the parallel null lines, $v =-A$ and $v=-A\a(V)$,
with boundary identified by a boost with velocity $V$ as above; and
take a finite section $S$ of that strip bounded by the bottom half of
the hyperbola $uv = A^2$ and by the left half of the hyperbola $uv =
-A^2$, as in Fig.~\ref{mtt} (Here $u=t-z,\ v=t+z$ are the usual
null coordinates).  Finally, take as the piece of 4-dimensional
Minkowski space $S\times D$, where $D$ is the disk $x^2+y^2 < B^2$, and
$B>A$.   Spacelike sections of $S\times D$ prior to the horizon, ${\cal
C}\times D$, are cylinders with circular cross section in $E^2$.  In
the full spacetime, $\cN$, the surface ${\cal C}\times D$ is again a
part of the chronology horizon.

The key to showing that the chronology horizon of $\cN$ is unstable is
to note that the past of points on $C\times D$ with $x=y=0$ lies
entirely in $C\times D$.  This is most easily seen using the universal
covering space of $S\times D$.  Misner space has as its universal
covering space the half of 2-dimensional Minkowski space to the left
of $v=0$.  The corresponding cover $\bar S\times D$ of $S\times D$ is
the part of 4-dimensional Minkowski space bounded by the 3-surfaces
$v=-A$; $v=-A\a$; $uv=A$, lower branch; $uv=-A$, left branch; and
$x^2+y^2<B^2$.  The past light cone in $\bar S\times D$ of a point on
$\bar{\cal C}\times (0,0)$ has maximum value of $x^2+y^2$ where it
meets the boundary $uv=A$, and calculation shows that the intersection is a
surface with
$x^2+y^2 < A^2$.  Since, by construction, $A^2 < B^2$, the past light
cone of a point of $\bar{\cal C}\times\{(0,0)\}$ never intersects the
boundary $x^2+y^2 = B^2$.  Thus every point $\bar P\in\bar S\times D$
to the past of $\bar{\cal C}\times \{(0,0)\}$ is in the domain of
dependence of the spacelike boundary $uv=A$ of $\bar S\times D$; and
every point $P\in S\times D$ to the past of $ {\cal C}\times (0,0) $ is
then in the domain of dependence of the
boundary $uv=A$ of $S\times D$.  Data on $uv=A$ that is independent of
$x$ and $y$ yields a solution that diverges in $S\times D$ because the
solution is identical in the domain of dependence of $uv=A$ to the
divergent solution in Misner space.  Finally, by picking a spacelike
hypersurface of the full spacetime that agrees with the $uv=A$ surface
in $S\times D$, we obtain data on a spacelike hypersurface to the past
of the chronology horizon for which no finite-energy solution exists to the
scalar wave equation for generic smooth initial data with finite
energy.

The example of a geometry with compact nonchronal region for which
{\it uniqueness} fails depends on a construction suggested by
Geroch\cite{gerochpvt}.
Although 2-dimensional geometries obtained by removing slits and identifying
sides are singular \cite{cgs}, it is
possible in 4-dimensions to build smooth
geometries in a similar way. The construction relies on the following
observation.

\noindent
{\sl Lemma 9} Any smooth compact 4-dimensional spacetime with boundary can be
embedded\\
(i) in a smooth compact spacetime without boundary, and\\
(ii) in a smooth spacetime that is isometric to Minkowski space outside a
compact region.\\

\noindent
{\sl Proof.} The technique is borrowed from references
\cite{geroch,sorkin} (see also \cite{f91}).
(i): Any manifold $M$ with boundary $\Sigma$ can be embedded in
a compact manifold $\tilde M$ by attaching a second copy $M'$ of $M$ to
the outward side of $\Sigma$. Let $U$ be a collar of $M$, a
neighborhood $U\cong\Sigma\times I$ with boundary
$\Sigma'\sqcup\Sigma$.  A lorentzian metric $g$ on $M$ can be
extended to a lorentzian metric $\tilde g$ on $\tilde M$ precisely when a
direction field $\hat t$ of timelike directions on $U$ can be extended
to a timelike direction field on $\tilde M$. Now $\hat t$ can always be
extended to a Morse direction field on $\tilde M$, a direction field that has
isolated zeroes, at each of which the line-element field is tangent to
a vector field with index $\pm 1$.  By cutting out a ball $B^4$
containing each zero and gluing in a copy of $\sRR P^4\backslash B^4$
for each zero of index $1$ and a copy of $\CC P^2\backslash B^4$ for
each zero of index $-1$, we can extend the line element to a
nonvanishing field on the interior. \\
(ii): The proof here is nearly identical. First embed $M$ in $\tilde M $ as in
(i). If one removes a a ball
$B^4$ from $ M'$ then one can put a flat metric on a
neighborhood $V$ of the spherical boundary $\partial B^4$ that makes
$V$ into a copy of Minkowski space outside a ball.  One is again asking
to extend a direction field on a new boundary, $U\sqcup V$ to the
4-manifold that it bounds, and the construction proceeds as in (i).
Finally any Lorentzian metric $g$ on the compact
manifold-with-boundary, $M'$ ( $M'\backslash B^4$) that is smooth on
$U$ ($U\sqcup V$) can be deformed to a smooth metric that agrees with
$g$ on a neighborhood of the boundary of $U$ and $V$. $\qquad\qquad\square$\\

Using this construction, we exhibit a smooth, asymptotically flat
spacetime with compact nonchronal region, for which solutions with
finite energy do not exist for generic initial data. Begin with a
4-torus $T^4$ with a flat metric $\eta$ chosen to make two of the
generators null and the other two spacelike. Explicitly, identify by
translation opposite faces of the rectangular 4-cell in Minkowski
space, $0 \leq u \leq A,\ 0\leq v \leq A, 0 \leq x\leq A,\ 0 \leq y\leq
A.$ The geometry $T^4,\ \eta$ is chosen because it has solutions to the
wave equation whose support is not all of $T^4$; examples are smooth
plane waves, functions $\Phi = \Phi(u),$ with $\Phi(u)=0, u<A/2$.  Cut
a ball out of $T^4$ and embed it in a spacetime that is isometric to
Minkowski space outside a compact region $U$, using {\sl Lemma 9}.
\footnote{An example of suitable manifold is $T^4 \# R^4 \# CP^2\#
CP^2$.}  The resulting spacetime satisfies condition (i) of the {\sl
Conjecture}, but solutions to the wave equation for data on a Cauchy
surface $\cM$ for the past of the Cauchy horizon are not unique:  Zero
data on $\cM$ is consistent with arbitrary solutions $\Phi(u)$ whose
support on $T^4$ is disjoint from the removed ball.\\

Although our primary interest is in smooth geometries with CTCs
confined to a compact region, it is worth pointing out that if one
allows singularities,  there are simple examples of spacetimes with
CTCs for which one can easily prove the existence of
solutions to free-field equations for arbitrary initial data.  For
these geometries, however, solutions for smooth data with finite energy
are not smooth and do not in general have finite energy; generic
solutions are in $L_2^{\rm loc}$.  Consider, for example,
two-dimensional Minkowski space with two parallel timelike or spacelike
segments of equal length removed, as in Fig.~\ref{slits}, and each side
of each segment identified with a side of the other segment after
translation by a timelike vector $V$, which will be taken to
point up and to the right.\footnote
{More precisely, to construct the first spacetime, let $\widehat\cN$ be
the manifold obtained from Minkowski space by removing the two timelike
segments $\bar L_1$ and $\bar L_2 = \cT (\bar L_I)$ where $\cT$ is
translation by a timelike vector $V$.  Call the segments with their
endpoints removed $L_1$ and $L_2$.  Formally reattach $L_1$ and $L_2$
by writing $\cN=\widehat\cN\sqcup L_1\sqcup L_2$.  The topology of the
first spacetime, $\cN$, is generated by the open sets of $\widehat\cN$
together with neighborhoods of points of $L_1$ and $L_2$ defined as
follows: Let ${\cal O}$ be any open set in an atlas for Minkowski space
intersecting the line through $\bar L_1$ in an open interval
$\ell\subset L_1$. Let ${\cal O}_R$ be the part of ${\cal O}$ lying to
the left (right) of the line through $L_1$.  Let ${\cal O'}_R $
be the part of the translated open set ${\cal O'}=\cT({\cal O})$ that
lies to the right of the line through $L_2$; and let ${\cal O'}_L
$ be the part of the translated open set ${\cal O'}=\cT({\cal O})$ that
lies to the left of the line through $L_2$ and to the right of the
line through $\bar L_1$.  Then ${\cal O}_L\cup \ell \cup
{\cal O}_R$ and ${\cal O'}_R\cup \ell' \cup {\cal O'}_L$ are open sets of
$\widehat\cN$.  With the obvious maps to subsets of $\sRR^4$, the open
sets just enumerated form an atlas.  Because of the deleted endpoints,
$\cN$ is not complete.}

The resulting geometries are flat with two conical singularities,
corresponding to the removed endpoints of the segments.  Similar
spacetimes have been discussed by Geroch and Horowitz\cite{gh} and
Politzer\cite{politzer}.  Some identified points are timelike related, and
timelike curves in Minkowski space joining points $P_I \in L_I$ to
$\cT(P_I)\in L_{II}$ become CTCs in $\cN$.  For each spacetime $\cN$
there is a spacelike hypersurface $\cM$ to the past of the nonchronal
region on which initial data can be set, and we will see it is easy to
find a solution for arbitrary initial data on $\cM$.

\noindent
{\sl Proposition 4} For any initial data $\phi, \dot \phi$ in
$L_2(\cM)\otimes L_2(\cM)$ there is a unique solution in $L_2^{\rm
loc}$ to the massless scalar wave equation on spacetimes of the form
described above.  In a neighborhood of spatial infinity the solution
agrees with the Minkowski space solution for the same initial data.

\noindent
{\sl Proof}.  We can divide initial data on $\cM$ into a sum
of data for right-moving and left-moving waves, $f(x-t)$ and $g(x+t)$,
by writing
\beq
f(x) = {1\over2}[\phi(x) - \int_{-\infty}^x dx' \dot\phi(x')],\qquad
g(x) = {1\over2}[\phi(x) + \int_{-\infty}^x dx' \dot\phi(x')].
\eeq
We separately construct solutions for right-moving and left moving
data.  On Minkowski space, right-moving data, $(f, \dot f = -f')$ gives
the solution $f(x-t)$; equivalently, $f(P) = f(p)$, where $p\in\cM$ is the
past endpoint of the right moving null ray from $\cM$ to $P$.  Note
that data in $L_2$ that is discontinuous across a finite set of points
$p, q, \cdots r, s$ of $\cM$ yields a solution that is discontinuous
across the boundaries of the strip between the two right-moving null
lines through endpoints $p$, $q$.  Each point of $\cN$ similarly lies
on a unique right-moving null geodesic, and all but four of these rays,
followed back to the past, intersect $\cM$.  Define a solution in
$L_2^{\rm loc}(\cN)$ by $f(P)=f(p)$, where $p\in\cM$ is the past
endpoint of the right moving null ray from $\cM$ to $P$. The four rays
that fail to meet $\cM$ are the future parts of null lines that emerge
from the (removed) endpoints of the identified segments.  These lines
are the future and past parts of right-moving null rays that are
geodesically incomplete, leaving the manifold at the removed
endpoints.  They divide $\cN$ into five strips which intersect $\cM$ in
five segments
$$
(-\infty,{p_2}'),\ ({p_2}',p_2),\ (p_2, {p_1}'),\ ({p_1}', p_1),\ (p_1,
\infty),
$$
where $p_i$ and ${p_i}'$ are the points where the right-moving null
rays from the bottom and top endpoints of $L_i$, respectively, meet
$\cM$.  Since each strip is isometric to a strip bounded by null rays in
Minkowski space, the function $f$ satisfies the scalar wave equation
with the given initial data everywhere except at the null boundaries of
the strips.

The proof of existence for left-moving solutions is identical,
where five new strips intersect $\cM$ in five new segments
$$
(-\infty,q_1),\ (q_1,{q_1}'),\ ({q_1}',q_2),\ (q_2,{q_2}'),\ ({q_2}',\infty),
$$
where $q_i$ and ${q_i}'$ are the points where left-moving null rays
from the bottom and top endpoints of $L_i$, respectively, meet
$\cM$.
Outside the chronology horizon, the left- and right-moving
solutions have their Minkowski space values because past directed null
rays from points outside the horizon never intersect $L_1\cup L_2$.
Thus the solution outside a spatially compact region of any
asymptotically spacelike hypersurface agrees with the Minkowski space
solution for the same initial data.  $\qquad\qquad\square$

Proving uniqueness of these solutions appears to be straightforward:
Assuming that any solution in $L_2^{\rm loc}$ is locally a sum of
right-moving and left-moving solutions, one can trace it back to
nonzero data on $\cM$.  Similar spacetimes can be constructed in
4-dimensions by removing two parallel planar 3-disks from Minkowski space
and identifying their boundaries as in the two dimensional example.
Again it seems clear that solutions in $L_2^{\rm loc}$ exist and that
they do not in general have finite energy.

Curiously, in 4-dimensions these singular spacetimes can be made into
smooth spacetimes by using a construction essentially equivalent to
that of {\sl Lemma 9}.  In addition to removing two 3-disks, one
removes a small solid torus ($D^3\times S^1$) at the edge of each disk.
Then when the sides of the disk are glued back in, one is left with a
spacetime with boundary $S^3\times S^1$, which one can glue to a
compact spacetime.

\acknowledgments
We thank Piotr Chrusciel, Robert Geroch, and Robert Wald for helpful
discussions and Rainer Picard for extensive coaching.  M.S. Morris
wishes to acknowledge support from a Holcomb Research Fellowship at
Butler University while completing this work.  The work was also
supported by NSF Grant No. PHY-9105935.

\begin{figure}
\caption{An orientable 3-manifold $M$ is constructed by identifying points of
$\Sigma_I$ and points of $\Sigma_{II}$ that are labeled by the same letter,
with subscripts I and II: $P_{II} = \cT(P_I)$.}
\label{M}
\end{figure}

\begin{figure}
\caption{The spacelike hypersurfaces $\cM_t$ foliate the spacetime
whose boundary is the union of two cylinders, $C_I\cup C_{II}$. In
the spacetime $\cN$, the cylindrical boundaries are identified, and
$\cM_{t+\tau}$ is a smooth continuation of $\cM_t$ across the
identified spheres $\Sigma_I$ and $\cT(\Sigma_I)$.}
\label{Mt}
\end{figure}

\begin{figure}
\caption{Misner space is the region between the two null rays $\ell$
and $\cB\ell$, with points of the null boundaries identified by the boost
$\cB$.  The curve ${\cal C} = CC'$ is a chronology horizon, a closed null
geodesic that separates the nonchronal region above it from the
globally hyperbolic spacetime to its past. Equivalently, Misner space
is the quotient of the half of Minkowski space lying to the left of the
null line L by the group of boosts generated by $\cB$.}
\label{misner}
\end{figure}

\begin{figure}
\caption{A piece of Misner space is used in the construction of a
4-dimensional spacetime $\cN$ with a compact nonchronal region and an
unstable chronology horizon. The lines pq and p'q' lie in the covering
space, the half of Minkowski space lying to the left of $v=0$, and they
are identified by the boost that defines Misner space.  In $\cN$, these
lines can be regarded as lying along the trajectory of the wormhole
mouth.}
\label{mtt}
\end{figure}

\begin{figure}
\caption{Two simple spacetimes with CTCs and a well defined Cauchy
problem are shown these two figures.  Two parallel slits are removed
from Minkowski space and points labelled by the same letter are
identified. }
\label{slits}
\end{figure}

\begin{figure}
\caption{Each shaded region is a strip isometric to a piece of Minkowski
space.  A right-moving solution to the massless wave equation is smooth
and well-defined on each strip.}
\label{regions}
\end{figure}

\begin{figure}
\caption{An example of an asymptotically flat spacetime with compact
nonchronal region is depicted here with two dimensions suppressed.
Balls are removed from ${\cal N}$ and the torus, and their boundary
3-spheres $\Sigma_I$ and $\Sigma_{II}$ are identified.  Arrows at $P$
point along null generators of the torus.  The shaded region is the
support of a solution to the massless scalar wave equation.}
\label{counterex}\end{figure}

\end{document}